\documentclass[12pt]{article}

\usepackage[totalwidth=467truept,totalheight=620truept]{geometry}
\usepackage{amsfonts,latexsym,amssymb,amsmath,graphicx,accents,eucal,slashed,subfigure}
\usepackage{dsfont}
\usepackage[T1]{fontenc}
\usepackage[hidelinks]{hyperref}

\newtheorem{definition}{Definition}
\newtheorem{theorem}{Theorem}
\newtheorem{lemma}{Lemma}
\newtheorem{proposition}{Proposition}

\global\arraycolsep=1truept

\numberwithin{equation}{section}

\begin{document}

\bigskip \phantom{C}

\vskip2truecm

\begin{center}
{\huge \textbf{Algebraic Cutting Equations}}

\vskip2truecm

\textsl{Damiano Anselmi}

\vskip .2truecm

\textit{Dipartimento di Fisica ``Enrico Fermi'', Universit\`{a} di Pisa, }

\textit{Largo B. Pontecorvo 3, 56127 Pisa, Italy}

\textit{and INFN, Sezione di Pisa,}

\textit{Largo B. Pontecorvo 3, 56127 Pisa, Italy}

damiano.anselmi@unipi.it

\vskip2.5truecm

\textbf{Abstract}
\end{center}

The cutting equations are diagrammatic identities that are used to prove
perturbative unitarity in quantum field theory. In this paper, we derive
algebraic, upgraded versions of them. Differently from the diagrammatic
versions, the algebraic identities also holds for propagators with
arbitrary, nonvanishing widths. In particular, the cut propagators do not
need to vanish off shell. The new approach provides a framework to address
unsolved problems of perturbative quantum field theory and a tool to
investigate perturbative unitarity in higher-derivative theories that are
relevant to the problem of quantum gravity, such as the Lee-Wick models and
the fakeon models.

\vfill\eject

\section{Introduction}

\label{s0}

\setcounter{equation}{0}

Perturbative unitarity in quantum field theory is the statement that the
scattering matrix $S$ is unitary at the perturbative level. This property
can be rephrased as a set of identities obeyed by the scattering amplitudes.
At the diagrammatic level, it amounts to a set of \textit{cutting equations }%
\cite{cutkosky}, which involve a diagram $G$ together with the variants
obtained by cutting $G$ in various ways. In this paper, we show that
perturbative unitarity can be conceptually reduced to a set of polynomial
equations, which we call \textit{algebraic cutting equations}. They are
actually more general than what is strictly needed for the proof of
perturbative unitarity, which is why we think that they deserve
consideration on their own as mathematical properties.

The usual proof of perturbative unitarity \cite{cutkosky,thooft,brsunitarity}
proceeds in four steps, which are the derivations of: ($i$) a diagrammatic
equation in coordinate space known as \textit{the largest time equation}, ($%
ii$) the cutting equations properly known (henceforth called \textit{%
diagrammatic} cutting equations, to distinguish them from the algebraic
ones), ($iii$) the pseudounitarity equation, and ($iv$) the unitarity
equation $SS^{\dagger }=1$. In this paper, we concentrate on a new,
algebraic approach that allows us to jump directly to point ($ii$). We will
not have much to say about the other steps just mentioned, which remain
unmodified.

In particular, the projection ($iii$) $\rightarrow $ ($iv$) of the
pseudounitarity equation onto the unitarity equation is necessary only in
the presence of local symmetries. Its role is to show that the temporal and
longitudinal components of the gauge fields are compensated by the
Faddeev-Popov ghosts. The projection can be handled with the methods of
refs. \cite{thooft} and \cite{brsunitarity} in gauge theories and those of
ref. \cite{unitarity} in gauge theories and gravity. The implication ($ii$) $%
\rightarrow $ ($iii$) is straightforward, since the pseudounitarity equation
is just a collection of the cutting equations. The implication ($i$) $%
\rightarrow $ ($ii$) follows from the Fourier transform of the largest time
equation plus the requirement that positive energies propagate forward in
time and negative energies propagate backward in time. In the approach we
offer here, it is a particular case of the general theorem we prove.

In this paper, we replace the step ($i$) with a set of algebraic identities
that allow us to gain a deeper understanding into the mathematical aspects
of perturbative unitarity, and make various manipulations more efficiently.
Moreover, the algebraic cutting equations are more general than the usual
ones. Indeed, they also hold for arbitrary, nonvanishing widths, since the
cut propagators do not need to be distributions that are supported only on
shell.

The basic concept we need to build the identities is the concept of \textit{%
polar} number, which is a variable equipped with a \textit{polarity}. By
convention, the polarities are denoted by $+$ and $-$. The polarity is an
abstract marking that allows us to divide the set of variables we use into
two subsets: the subset made of the variables with positive polarity and the
subset made of the variables with negative polarity. In typical
applications, the polar numbers have complex values and the
positive/negative polarity can denote their locations inside/outside some
closed curve $\gamma $ on the complex plane or the Riemann sphere. If the
polar number is a function of another variable (typically an energy $E$) and
has a singularity for a specific value of $E$, then the polarity may refer
to the location of the singularity inside/outside a closed curve $\gamma $.

Given an oriented Feynman diagram $G$, we give rules to associate one polar
number with each internal leg. The product of such polar numbers is called
polar monomial. A polarized monomial is a polar monomial where at least one
loop is polarized, that is to say each leg of the loop is associated with a
polar number whose polarity agrees with the leg orientation. The theorem we
prove states that certain polynomials of polar numbers are equal to sums of
polarized monomials.

In the applications to physics, the legs of the diagrams are oriented
according to energy flows. The polar numbers are \textquotedblleft half
propagators\textquotedblright\ (the propagator being the sum of two polar
numbers). They depend on a momentum and have a pole for some complex value
of the energy. The polarity is positive or negative according to whether the
pole is located below or above the real axis of the complex energy plane.
The theorem singles out the polarized monomials, which do not contribute to
the diagrammatic cutting equations. The reason is that polarized loops give
zero when they are integrated on the loop momentum. The algebraic identities
thus lead to the diagrammatic cutting equations in a straightforward way.

The approach of this paper offers a clearer understanding of perturbative
unitarity, by uncovering its purely algebraic aspects. As we show in section %
\ref{conclusions}, it also helps organizing computations in more practical
ways. Moreover, the generalized versions that hold for arbitrary widths
allow us to upgrade the formulation of unitarity to include the effects of
radiative corrections, which typically generate nonvanishing widths at one
and higher loops. Several aspects of this inclusion have yet to be clarified 
\cite{diagrammar}.

Finally, the algebraic cutting equations provide the best framework to
investigate perturbative unitarity in theories that have not been reached so
far by the standard techniques. Examples are the Lee-Wick models \cite%
{leewick}, which do involve propagators with nonvanishing widths. They are
higher-derivative theories of a special class that are claimed to reconcile
renormalizability with unitarity. The Lee-Wick models have been studied in a
variety of contexts \cite{contexts} and are expected to have important
implications for quantum gravity \cite{LWgrav,LWgravmio}. They have been
reformulated as nonanalytically Wick rotated Euclidean theories in ref. \cite%
{LWformulation} and their unitarity has been proved at one loop in ref. \cite%
{piva}. They admit important generalizations where the would-be ghosts are
turned into \textquotedblleft fakeons\textquotedblright , i.e. fake degrees
of freedom, by means of a new quantization prescription \cite{LWgravmio}.
Using the algebraic cutting equations, a proof of unitarity to all orders
has been recently provided in ref. \cite{fakeons} for all the theories that
contain fakeons and physical degrees of freedom.

The paper is organized as follows. In section \ref{s1} we collect the basic
definitions. In section \ref{s2} we state the main theorem, which we prove
in section \ref{s3}. In section \ref{examples} we give a number of examples.
Specifically, we use the algebraic identities to derive the diagrammatic
cutting equations of the bubble and triangle diagrams at one loop and the
chestnut diagram at two loops. We include the algebraic identities of other
diagrams, up to three loops. In section \ref{s5} we use the identities to
prove the perturbative unitarity of ordinary quantum field theories. In
section \ref{paritytr} we discuss some symmetries of the algebraic cutting
equations. Section \ref{conclusions} contains the conclusions, with emphasis
on the virtues of the algebraic approach with respect to the usual approach.

\section{Basic definitions}

\setcounter{equation}{0}\label{s1}

In this section we collect the basic definitions that are necessary to state
the main theorem.

A diagram is a set of vertices connected by lines. The lines of a diagram
will be called \textit{legs} henceforth. The diagrams we consider do not
need to be planar or connected. The vertices can be the endpoints of any
number of legs, including one or two. The vertices that are attached to a
unique leg are called external. The legs they are attached to are also
called external. The other vertices and legs are called internal. From now
on, we drop the external vertices and whenever we talk about vertices we
mean the internal ones.

Equip the internal legs of the diagrams with orientations. The definition of
oriented leg is self evident. Two legs are called adjacent if they have a
vertex in common. Two adjacent legs are said to have coherent orientations
if the orientation of one leg points to the vertex in common and the
orientation of the other leg points away from the vertex in common.

\begin{definition}
Given a diagram, a curve is a sequence $\{\ell _{1},\ldots \ell _{n}\}$ of
legs $\ell _{i}$, such that each $\ell _{i}$ with $i>1$ is adjacent to $\ell
_{i-1}$. A loop is a closed curve, i.e. a curve $\{\ell _{1},\ldots \ell
_{n}\}$ such that $\ell _{1}$ is adjacent to $\ell _{n}$. A curve is minimal
if it contains no loop. A loop is minimal if it contains no loop apart from
itself.
\end{definition}

An example of nonminimal loop is a loop that looks like an \textquotedblleft
8\textquotedblright .

\begin{definition}
A curve or a loop are oriented if the orientations of all their legs are
coherent.
\end{definition}

Assume that $G$ is connected and has $I$ internal legs and $V$ vertices.
Pick $I$ independent real numbers $E_{i}$, $i=1,\ldots I$, and call them
\textquotedblleft energies\textquotedblright . Assign an energy to each
internal leg and zero energy to each external leg. Use the orientation of a
leg to define the orientation of the flow of its energy. Then, impose the
energy conservation at each vertex. This is the requirement that the total
energy flowing into the vertex must be equal to the total energy flowing out
of the vertex. The independent conservation conditions are $V-1$, because
the energies flowing into the diagram and out of it are zero by assumption.
Due to this, the energy is automatically conserved in the last vertex, once
it is conserved in every other vertex. Energy conservation leaves us with $%
I-V+1\equiv L$ arbitrary independent energies $e_{1},\ldots e_{L}$.

\begin{proposition}
It is possible to arrange the leg orientations and the energies $%
e_{1},\ldots e_{L}$, so that the flow of each energy defines an oriented
minimal loop in $G$ and each leg is associated with a linear combination of
energies $e_{1},\ldots e_{L}$ with coefficients 0 or 1. \label{p1}
\end{proposition}

\textbf{Proof}. To see this, start from the diagram $G$, with no leg
orientations and zero energy in every leg. Assume, for the time being, that $%
G$ is one-particle irreducible. Consider a minimal loop $\gamma _{1}$ in $G$%
. Arrange the orientations of the $\gamma _{1}$ legs so that they are
coherent and add the energy $e_{1}$ to each of its legs. So doing, the loop $%
\gamma _{1}$ becomes oriented. If $L=1$, the construction stops here.

Otherwise, since $G$ is one-particle irreducible, there must exist a pair $%
v^{(a)}$, $v^{(b)}$ of $\gamma _{1}$ vertices that are connected by a
minimal curve $\Gamma _{ab}$ which has no legs and no other vertex in common
with $\gamma _{1}$. We distinguish two cases: $v^{(a)}\neq v^{(b)}$ and $%
v^{(a)}=v^{(b)}$. If $v^{(a)}\neq v^{(b)}$, $v^{(a)}$ and $v^{(a)}$ are
connected both by $\Gamma _{ab}$ and by two portions $\Delta _{ab}$ and $%
\Delta _{ab}^{\prime }$ of $\gamma _{1}$. Pick the portion of your choice,
say $\Delta _{ab}$. The union $\Gamma _{ab}\cup \Delta _{ab}$ defines a
minimal loop $\gamma _{2}$. The orientation of $\Delta _{ab}$ can be
extended coherently to $\Gamma _{ab}$, to define the orientation of $\gamma
_{2}$. Once this is done, add the energy $e_{2}$ to each leg of $\gamma _{2}$%
. If $v^{(a)}=v^{(b)}$, just pick $\Gamma _{ab}$ as the loop $\gamma _{2}$
and orient it in the way you like. Then add $e_{2}$ to each of its legs. If $%
L=2$, the construction stops here.

Observe that any distinct vertices $v^{(c)}$ and $v^{(d)}$ of $\gamma
_{1}\cup \gamma _{2}$ are connected by an oriented minimal curve $\Delta
_{cd}$ contained in $\gamma _{1}\cup \gamma _{2}$: if they both belong to $%
\gamma _{1}$ or $\gamma _{2}$, this fact is obvious. If $v^{(c)}$ belongs to 
$\gamma _{1}$ and $v^{(d)}$ belongs to $\gamma _{2}$, it is sufficient to
move along $\gamma _{1}$ (following the $\gamma _{1}$ orientation) from $%
v^{(c)}$ to the first intersection between $\gamma _{1}$ and $\gamma _{2}$,
then continue to $v^{(d)}$ along the portion of $\gamma _{2}$ that has a
coherent orientation. Clearly, such a $\Delta _{cd}$ is a minimal curve.

If $L>2$, there must exist a pair of vertices $v^{(c)}$ and $v^{(d)}$ of $%
\gamma _{1}\cup \gamma _{2}$ that are connected by a minimal curve $\Gamma
_{cd}$ that has no legs and no other vertex in common with $\gamma _{1}\cup
\gamma _{2}$. If $v^{(c)}\neq v^{(d)}$, by the property shown above they are
also connected by an oriented minimal curve $\Delta _{cd}$ contained in $%
\gamma _{1}\cup \gamma _{2}$. The union $\Gamma _{cd}\cup \Delta _{cd}$ of
the two curves defines the third minimal loop $\gamma _{3}$, which becomes
oriented after the orientation of $\Delta _{ab}$ is coherently extended to
the whole loop. Finally, the energy $e_{3}$ is added to all the legs of $%
\gamma _{3}$. If $v^{(c)}=v^{(d)}$, just pick $\Gamma _{cd}$ as $\gamma _{3}$%
, orient it in the way you like and add $e_{3}$ to each of its legs. If $L=3$%
, the construction stops here.

Again, any pair of distinct vertices that belong to the union $\gamma
_{1}\cup \gamma _{2}\cup \gamma _{3}$ are connected by an oriented curve
contained in $\gamma _{1}\cup \gamma _{2}\cup \gamma _{3}$, which we can
choose to be minimal. This allows us to iterate the construction for $L>2$.

It is also straightforward to extend the assignments to the one-particle
reducible diagrams as well as the disconnected diagrams. This concludes the
proof. $\square $

\begin{definition}
A diagram is oriented if its leg orientations are compatible with the
construction of proposition \ref{p1}. Otherwise, the diagram is jammed.
\end{definition}

\begin{figure}[t]
\begin{center}
\includegraphics[width=10truecm]{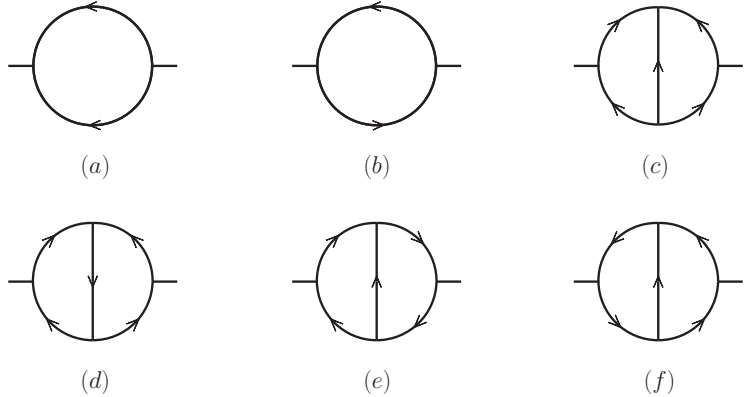}
\end{center}
\caption{Orientations of diagrams}
\label{orientation}
\end{figure}

We call $e_{1},\ldots e_{L}$ \textit{loop energies}. A diagram may admit
various orientations, which give equivalent polynomial identities. In fig. %
\ref{orientation} we show various examples. The diagrams ($a$) and ($c$) are
jammed, while ($b$), ($d$), ($e$) and ($f$) are oriented. In particular, ($d$%
), ($e$) and ($f$) are different orientations of the same diagram. A diagram
has precisely $L$ oriented minimal loops. For example, ($d$), ($e$) and ($f$%
) have two oriented minimal loops.

\section{The theorem}

\setcounter{equation}{0}\label{s2}

In this section we state the theorem, leaving its proof to the next section.

Let $G$ denote an oriented diagram and $\ell _{1},\ldots \ell _{I}$ its
internal legs. Let $\nu _{a}$, $a=1,\ldots V$, label the vertices of the
diagram. When we want to specify that the orientation of the $i$th leg
points from, say, the vertex $\nu _{a}$ to the vertex $\nu _{b}$, we denote
it by $(\nu _{a}\ell _{i}\nu _{b})$.

Build variants $G_{M}$ of $G$ by marking any number of vertices. We denote
the marked vertices by $\hat{\nu}_{a}$. The marked diagrams have legs of
types $(\nu _{a}\ell _{i}\hat{\nu}_{b})$, $(\hat{\nu}_{a}\ell _{i}\nu _{b})$
and $(\hat{\nu}_{a}\ell _{i}\hat{\nu}_{b})$, besides those of type $(\nu
_{a}\ell _{i}\nu _{b})$, the leg orientation pointing from $a$ to $b$.

A \textit{polar} number is a variable equipped with a \textit{polarity},
denoted by $+$ or $-$. Let $\{\sigma _{i}^{+},\tau _{i}^{+},\sigma
_{i}^{-},\tau _{i}^{-}\}$, $i=1,\ldots I$, denote $I$ quartets of polar
numbers. Each quartet is associated with a leg of the diagram and is the
union of a pair $\sigma _{i}^{+}$, $\tau _{i}^{+}$ of variables with
positive polarities and a pair $\sigma _{i}^{-}$, $\tau _{i}^{-}$ of
variables with negative polarities.

\begin{figure}[t]
\begin{center}
\includegraphics[width=14truecm]{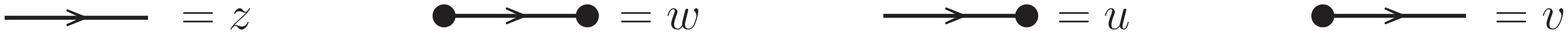}
\end{center}
\caption{Propagators}
\label{feybub}
\end{figure}

Define the \textit{propagators}%
\begin{equation}
z_{i}=\sigma _{i}^{+}+\sigma _{i}^{-},\qquad w_{i}=\tau _{i}^{+}+\tau
_{i}^{-},\qquad u_{i}=\sigma _{i}^{+}+\tau _{i}^{-},\qquad v_{i}=\sigma
_{i}^{-}+\tau _{i}^{+}.  \label{propagators}
\end{equation}%
Determine the \textit{value} $P_{M}$ of the diagram $G_{M}$ by means of the
following \textquotedblleft Feynman\textquotedblright\ rules. Assign the
value one to each unmarked vertex and the value $-1$ to each marked one.
Associate propagators with the legs of $G_{M}$ as follows: 
\begin{equation}
(\nu \ell _{i}\nu ^{\prime })\rightarrow z_{i},\qquad (\hat{\nu}\ell _{i}%
\hat{\nu}^{\prime })\rightarrow w_{i},\qquad (\nu \ell _{i}\hat{\nu}^{\prime
})\rightarrow u_{i},\qquad (\hat{\nu}\ell _{i}\nu ^{\prime })\rightarrow
v_{i}.\qquad  \label{scheme}
\end{equation}%
Graphically, we denote the marked vertices by means of a dot, so the
propagators are those shown in fig. \ref{feybub}. Then, $P_{M}$ is the
polynomial 
\begin{equation}
P_{M}=(-1)^{m}\prod\limits_{i=1}^{I}p_{Mi},  \label{energy}
\end{equation}%
where $p_{Mi}$ denotes the propagator of the $i$th leg $\ell _{i}$, assigned
according to the scheme (\ref{scheme}), and $m$ is the number of marked
vertices.

\begin{figure}[b]
\begin{center}
\includegraphics[width=14truecm]{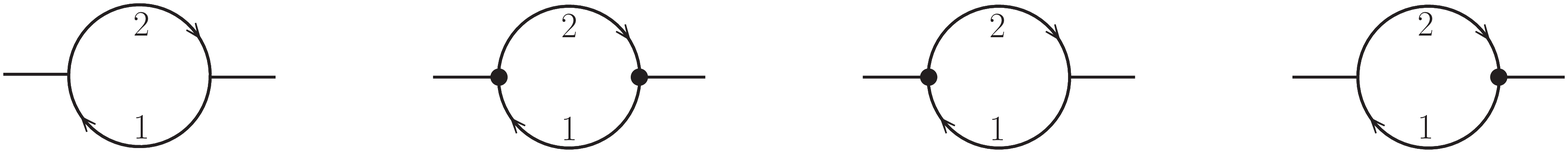}
\end{center}
\caption{Simple marked diagrams}
\label{polar}
\end{figure}

For example, the polynomials associated with the marked diagrams of fig. \ref%
{polar} are%
\begin{equation}
z_{1}z_{2},\qquad w_{1}w_{2},\qquad -u_{1}v_{2},\qquad -v_{1}u_{2},
\label{poll}
\end{equation}%
respectively.

The set or marked diagrams includes the diagram $G$ itself, where all the
vertices are unmarked, as well as the diagram $\bar{G}$ where all the
vertices are marked. The polynomials $P$ and $\bar{P}$ associated with $G$
and $\bar{G}$ are%
\begin{equation*}
P=\prod\nolimits_{i=1}^{I}z_{i},\qquad \bar{P}=(-1)^{V}\prod%
\nolimits_{i=1}^{I}w_{i},
\end{equation*}%
respectively.

The theorem is about the sum of the polynomials $P_{M}$ on all the ways $M$
to mark the diagram $G$. Writing $z$, $w$, $u$ and $v$ as sums of polar
numbers, according to formula (\ref{propagators}), we can expand the sum of $%
P_{M}$ as a sum of \textit{polar monomials}. A polar monomial is the product
of one polar number of the set $\{\sigma _{i}^{+},\tau _{i}^{+},\sigma
_{i}^{-},\tau _{i}^{-}\}$ for each leg $\ell _{i}$. A polar curve, loop or
diagram is a curve, loop or diagram whose legs are equipped with polar
numbers.

Examples of polar monomials for the diagram $G$ of fig. \ref{polar} are%
\begin{equation}
\sigma _{1}^{+}\sigma _{2}^{+},\qquad \sigma _{1}^{-}\tau _{2}^{-},\qquad
\sigma _{1}^{+}\sigma _{2}^{-},\qquad \sigma _{1}^{+}\tau _{2}^{-},\qquad
\sigma _{1}^{-}\tau _{2}^{+},  \label{pola}
\end{equation}%
etc.

\begin{definition}
A polarized loop is a polar loop where adjacent legs of coherent (opposite)
orientations carry polar numbers of coherent (opposite) polarities. \label%
{defpol}
\end{definition}

In particular, an oriented polar loop is polarized if all its legs carry
polar numbers of the same polarity. Instead, a polarized nonoriented loop is
such that the leg polarity flips if and only if the orientation flips.

\begin{definition}
A polarized monomial is a polar monomial, associated with a diagram G, where
at least one loop is polarized. \label{defpol2}
\end{definition}

\begin{figure}[t]
\begin{center}
\includegraphics[width=8truecm]{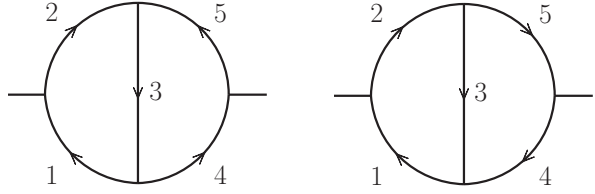}
\end{center}
\caption{Two-loop oriented diagrams}
\label{polloop}
\end{figure}

Consider, for example, the diagrams of fig. \ref{polloop}. The oriented
loops of the first diagram are 123 and 345, while 1254 is a nonoriented
loop. If we equip such loops with the polar monomials $\sigma _{1}^{+}\sigma
_{2}^{+}\tau _{3}^{+}$, $\sigma _{3}^{-}\tau _{4}^{-}\tau _{5}^{-}$ and $%
\sigma _{1}^{+}\sigma _{2}^{+}\sigma _{4}^{-}\tau _{5}^{-}$, respectively,
we obtain polarized loops. Examples of polarized monomials are $\sigma
_{1}^{+}\sigma _{2}^{+}\tau _{3}^{+}\sigma _{4}^{+}\tau _{5}^{-}$ and $%
\sigma _{1}^{+}\sigma _{2}^{+}\sigma _{3}^{-}\tau _{4}^{-}\tau _{5}^{-}$.
Examples of polarized loops for the second diagram of fig. \ref{polloop} are
123 with the monomial $\sigma _{1}^{+}\sigma _{2}^{+}\tau _{3}^{+}$, 1254
with $\sigma _{1}^{+}\sigma _{2}^{+}\tau _{4}^{+}\sigma _{5}^{+}$ and 345
with $\sigma _{3}^{+}\sigma _{4}^{-}\tau _{5}^{-}$.

Here is the main theorem of this paper:

\begin{theorem}
given a diagram $G$, the identity%
\begin{equation}
\sum_{\text{markings }M}P_{M}=\mathcal{P}_{G}  \label{theorem}
\end{equation}%
holds, where $\mathcal{P}_{G}$ is a sum of polarized monomials.
\end{theorem}

We can condense the theorem by saying that the sum of the marked diagrams is
equal to a sum of polarized diagrams.

For example, if we sum the polynomials (\ref{poll}), we can easily check the
identity%
\begin{equation}
z_{1}z_{2}+w_{1}w_{2}-u_{1}v_{2}-v_{1}u_{2}=(\sigma _{1}^{+}-\tau
_{1}^{+})(\sigma _{2}^{+}-\tau _{2}^{+})+(\sigma _{1}^{-}-\tau
_{1}^{-})(\sigma _{2}^{-}-\tau _{2}^{-}).  \label{deco}
\end{equation}%
Note that the right-hand side is a sum of polarized monomials. More examples
are given in section \ref{examples}.

At the tree level, we have $\mathcal{P}_{G}=0$. At one loop, we have the
general formula%
\begin{equation}
\mathcal{P}_{G}=\prod\nolimits_{i=1}^{I}(\sigma _{i}^{+}-\tau
_{i}^{+})+\prod\nolimits_{i=1}^{I}(\sigma _{i}^{-}-\tau _{i}^{-}),
\label{wou}
\end{equation}%
which we leave without proof, since it is not crucial for the rest of the
discussion.

We can assume that the diagram $G$ does not contain tadpoles, i.e. loops
made of a single leg that begins and ends at the same vertex. Indeed, if $G$
contains tadpoles, the theorem is trivial, since a tadpole is an oriented
loop and can obviously be written as the sum of two contributions, each
being a polar number, which is polarized by definition. Moreover, we can
also assume that $G$ is connected, since the theorem extends to disconnected
diagrams in an obvious way, once it is proved for connected diagrams.
Finally, we can assume $V>1$, since a diagram with a single vertex has no
internal leg (in which case the theorem is obvious) or is a tadpole.

\section{Proof of the theorem}

\setcounter{equation}{0}\label{s3}

In this section we prove the theorem. Since formula (\ref{theorem}) is a
polynomial identity, if we prove it for polar numbers belonging to open sets
of the complex plane, we automatically prove it for arbitrary polar numbers.
Thus, with no loss of generality, we can assume that the signs of the
imaginary parts of the polar numbers $\sigma _{i}^{+},\tau _{i}^{+},\sigma
_{i}^{-},\tau _{i}^{-}$ coincide with their polarities.

Let $G$ denote an oriented connected diagram with $I$ internal legs, $V$
vertices, $L=I-V+1$ loops and no tadpoles. Denote the internal legs by $\ell
_{i}$, $i=1,\ldots I$, and the vertices by $\nu _{a}$, $a=1,\ldots V$. We
can assume $V>1$ and equip $G$ with loop energies $e_{1},\ldots ,e_{L}$ in
the way specified by proposition \ref{p1}.

\begin{figure}[t]
\begin{center}
\includegraphics[width=3.5truecm]{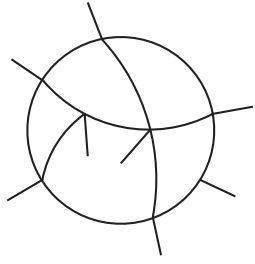}
\end{center}
\caption{Diagram with one external leg for each vertex}
\label{exte}
\end{figure}

If more external legs are attached to the same vertex, drop all of them but
one. If no external leg is attached to a vertex, add one. So doing, each
vertex is attached to one external leg and the diagram $G$ becomes a figure
like fig. \ref{exte}. Note that some external legs are drawn inside the
diagram, for no particular reason other than aesthetics. They could be
prolonged at will by crossing over the internal legs, since Feynman diagrams
need not be planar.

Assign propagators identically equal to one to the external legs. Then,
equip each external leg but the last one with an independent energy $%
\mathcal{E}_{a}$, $a=1,\ldots V-1$, that flows into the diagram. By the
conservation of energy, the external leg attached to the last vertex $\nu
_{V}$ has energy $\sum_{a=1}^{V-1}\mathcal{E}_{a}$ flowing out of the
diagram.

By the argument of section \ref{s1}, if $G$ is one-particle irreducible,
each pair of vertices $\{\nu _{a},\nu _{V}\}$, $a=1,\ldots V-1$, is
connected by a minimal curve $\Gamma _{a}$ in $G$ that is oriented from $\nu
_{a}$ to $\nu _{V}$. This property can be easily extended to any connected
diagram $G$. If $G$ is one-particle reducible, it can be viewed as a tree of
one-particle irreducible subdiagrams $G_{A}$, $A=1,\ldots N$, and single
(nonoriented) lines $\ell _{A}$, $A=1,\ldots N-1$, connecting pairs of $%
G_{A^{\prime }}$s. Equip $\ell _{A}$ with the orientation that flows towards
the subdiagram $\tilde{G}_{A}$ that contains $\nu _{V}$. Then, given a pair
of vertices $\{\nu _{a},\nu _{V}\}$, $a=1,\ldots V-1$, it is possible to
connect them through a minimal curve $\Gamma _{a}$ that is oriented from $%
\nu _{a}$ to $\nu _{V}$ and is equal to the union of a set of oriented
minimal curves $\Gamma _{A}^{\prime }\subset G_{A}$ and lines $\ell _{A}$.

Add the energy $\mathcal{E}_{a}$ to all the legs of $\Gamma _{a}$. Let $%
E_{i} $ denote the energy of the $i$th internal leg $\ell _{i}$. By
proposition \ref{p1} and the construction just described, $E_{i}$ is a
linear combination of the loop energies $e_{1},\ldots ,e_{L}$ and the
external energies $\mathcal{E}_{1},\ldots ,\mathcal{E}_{V-1}$, with
coefficients 0 or 1.

Define new polar numbers $\zeta _{i}^{\pm }$ and $\xi _{i}^{\pm }$ as%
\begin{equation}
\zeta _{i}^{\pm }=E_{i}-(\sigma _{i}^{\pm })^{-1},\qquad \xi _{i}^{\pm
}=E_{i}-(\tau _{i}^{\pm })^{-1}.  \label{enare}
\end{equation}%
Their polarities, specified by the superscripts $+$ and $-$, coincide with
the signs of their imaginary parts. We have%
\begin{equation}
\sigma _{i}^{\pm }(E_{i})=\frac{1}{E_{i}-\zeta _{i}^{\pm }},\qquad \tau
_{i}^{\pm }(E_{i})=\frac{1}{E_{i}-\xi _{i}^{\pm }}.  \label{enares}
\end{equation}%
For a while, we keep $\zeta _{i}^{\pm }$ and $\xi _{i}^{\pm }$ fixed and
treat $\sigma _{i}^{\pm }$ and $\tau _{i}^{\pm }$ as functions of the
energies. In some intermediate steps we integrate over the loop energies.
Then, we undo the integral to recover properties that hold at arbitrary
energies. This is the strategy that leads to the proof of the theorem.

Now, pick a marked diagram $G_{M}$, take formula (\ref{energy}), implement
the replacements (\ref{enares}) and integrate each loop energy $e_{j}$ along
the real axis with the measure $\mathrm{d}e_{j}/(2\pi )$. This defines the
value of $G_{M}$ in energy space, which is%
\begin{equation}
G_{M}(\mathcal{E}_{1},\ldots ,\mathcal{E}_{V-1})=(-1)^{m}\int
\prod\limits_{i=1}^{I}p_{Mi}(E_{i})\prod\nolimits_{j=1}^{L}\frac{\mathrm{d}%
e_{j}}{2\pi }.  \label{GM}
\end{equation}%
This integral is overall convergent, because it is well behaved at infinity
and no pole sits on the real axis. As far as the overall behavior at
infinity is concerned, observe that each polar number decreases like $1/e$,
where $e$ collectively denotes the loop energies, so the overall behavior of 
$P_{M}$ is $1/e^{I}$, which falls off fast enough, since $V>1$ implies $%
I=L+V-1\geqslant L+1$. Moreover, every subintegral is overall convergent for
a similar reason. Incidentally, the reason why we cannot treat diagrams that
contain tadpoles is that they do not satisfy these conditions.

We move to the coordinate versions of the diagrams, by taking their Fourier
transforms. The Fourier transforms of the polar numbers $\sigma _{j}^{\pm }$
and $\tau _{j}^{\pm }$ are%
\begin{equation*}
\tilde{\sigma}_{j}^{\pm }(t_{j})=\int_{-\infty }^{+\infty }\frac{\mathrm{d}%
E_{j}}{2\pi }\frac{\mathrm{e}^{iE_{j}t_{j}}}{E_{j}-\zeta _{j}^{\pm }}=\pm
i\theta (\pm t_{j})\mathrm{e}^{it_{j}\zeta _{j}^{\pm }},\qquad \tilde{\tau}%
_{j}^{\pm }(t_{j})=\pm i\theta (\pm t_{j})\mathrm{e}^{it_{j}\xi _{j}^{\pm }}.
\end{equation*}%
Familiar knowledge of quantum field theory tells us that the coordinate
version $\tilde{G}_{M}$ of (\ref{GM}) [multiplied by the distribution $(2\pi
)\delta (\mathcal{E}_{1}+\mathcal{E}_{2}+\cdots +\mathcal{E}_{V})$, which
imposes the overall energy conservation, where $\mathcal{E}_{V}$ is an
independent energy] is the product of the propagators in coordinate space,
times the values of the vertices, integrated over the locations $\mu _{a}$
of the vertices, i.e. 
\begin{equation*}
\tilde{G}_{M}(t_{1},\ldots ,t_{V})=(-1)^{m}\int \prod\limits_{a=1}^{V}\left[ 
\mathrm{d}\mu _{a}\hspace{0.01in}\delta (t_{a}-\mu _{a})\right]
\prod\limits_{j=1}^{I}\tilde{p}_{Mj}(\mu _{a_{j}}-\mu _{b_{j}}).
\end{equation*}%
Here $\tilde{p}_{Mj}(\mu _{a_{j}}-\mu _{b_{j}})$ denotes the Fourier
transform of the propagator $p_{Mj}$ associated with the line $\ell _{j}$
and $\mu _{a_{j}}$, $\mu _{b_{j}}$ are the time coordinates of the $\ell
_{j} $ endpoints, ordered so that the $\ell _{j}$ orientation points from
the vertex of time $\mu _{a_{j}}$ to the vertex of time $\mu _{b_{j}}$. The
delta functions are the Fourier transforms of the propagators of the
external legs (which are identically one in energy space).

The $\mu $ integrals are straightforward, so we just get%
\begin{equation*}
\tilde{G}_{M}(t_{1},\ldots ,t_{V})=(-1)^{m}\prod\limits_{j=1}^{I}\tilde{p}%
_{Mj}(t_{a_{j}}-t_{b_{j}}).
\end{equation*}

We can formulate the Feynman rules of the diagrams in coordinate space as
follows. As usual, the unmarked vertices are equal to one and the marked
vertices are equal to $-1$. The propagators $\tilde{p}_{Mj}(t_{a}-t_{b})$
are assigned according to the scheme 
\begin{eqnarray}
(\nu _{a}\ell _{j}\nu _{b}) &\rightarrow &z_{j}=\sigma _{j}^{+}+\sigma
_{j}^{-}\rightarrow i\theta (t_{ab})\mathrm{e}^{it_{ab}\zeta
_{j}^{+}}-i\theta (-t_{ab})\mathrm{e}^{it_{ab}\zeta _{j}^{-}},  \notag \\
(\nu _{a}\ell _{j}\hat{\nu}_{b}) &\rightarrow &u_{j}=\sigma _{j}^{+}+\tau
_{j}^{-}\rightarrow i\theta (t_{ab})\mathrm{e}^{it_{ab}\zeta
_{j}^{+}}-i\theta (-t_{ab})\mathrm{e}^{it_{ab}\xi _{j}^{-}},  \notag \\
(\hat{\nu}_{a}\ell _{j}\nu _{b}) &\rightarrow &v_{j}=\sigma _{j}^{-}+\tau
_{j}^{+}\rightarrow i\theta (t_{ab})\mathrm{e}^{it_{ab}\xi _{j}^{+}}-i\theta
(-t_{ab})\mathrm{e}^{it_{ab}\zeta _{j}^{-}},  \notag \\
(\hat{\nu}_{a}\ell _{j}\hat{\nu}_{b}) &\rightarrow &w_{j}=\tau _{j}^{+}+\tau
_{j}^{-}\rightarrow i\theta (t_{ab})\mathrm{e}^{it_{ab}\xi _{j}^{+}}-i\theta
(-t_{ab})\mathrm{e}^{it_{ab}\xi _{j}^{-}},  \label{sche}
\end{eqnarray}%
where $t_{ab}=t_{a}-t_{b}$ and $t_{a}$ denotes the time coordinate of the
vertex $\nu _{a}$.

Now we show that

\begin{lemma}
the identity%
\begin{equation}
\sum_{\text{markings }M}\tilde{G}_{M}(t_{1},\ldots ,t_{V})=0  \label{zero}
\end{equation}%
holds.
\end{lemma}

\begin{figure}[t]
\begin{center}
\includegraphics[width=10truecm]{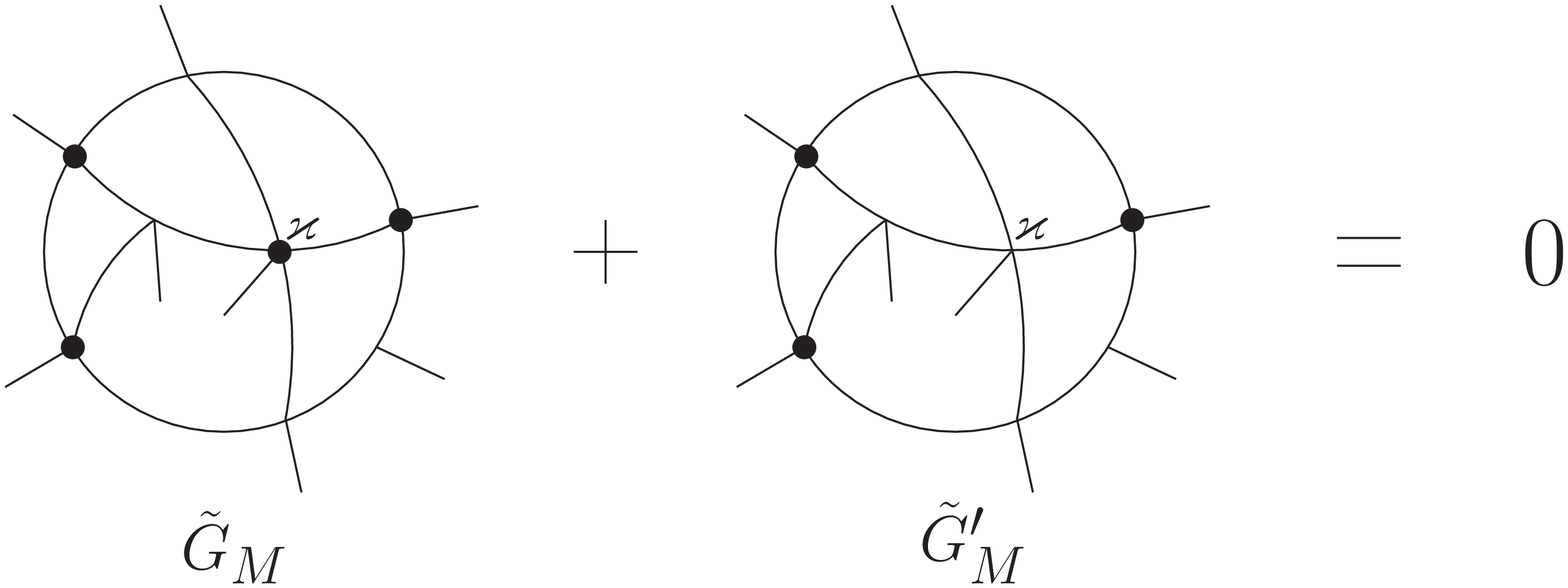}
\end{center}
\caption{Mutually canceling contributions}
\label{lowest}
\end{figure}

\textbf{Proof}. We can assume that the vertices have distinct times, because
the distributions (\ref{sche}) and the left-hand side of (\ref{zero}) do not
involve contact terms. Then, there is a lowest time, which we denote by $%
t_{0}$. Let $\varkappa $ denote its vertex. Formula (\ref{zero}) holds,
because the sum on the left-hand side contains pairs of mutually canceling
contributions, as shown in fig. \ref{lowest}. Specifically, for every
diagram $G_{M}$ that has $\varkappa $ unmarked, there is an almost identical
diagram $G_{M}^{\prime }$ that differs from $G_{M}$ just for the marking of $%
\varkappa $. The contribution $\tilde{G}_{M}^{\prime }$ due to $%
G_{M}^{\prime }$ is opposite to the contribution $\tilde{G}_{M}$ due to $%
G_{M}$, because the extra marking implies an extra minus sign. On the other
hand, all the propagators of $G_{M}$ and $G_{M}^{\prime }$ have equal
values. Those that do not involve $\varkappa $ coincide, because they
connect the same vertices. Those that involve $\varkappa $ coincide, because
the scheme (\ref{sche}) implies%
\begin{eqnarray*}
(\varkappa \ell _{j}\nu _{b}) &\rightarrow &-i\mathrm{e}^{i(t_{0}-t_{b})%
\zeta _{j}^{-}},\qquad (\varkappa \ell _{j}\hat{\nu}_{b})\rightarrow -i%
\mathrm{e}^{i(t_{0}-t_{b})\xi _{j}^{-}}, \\
(\hat{\varkappa}\ell _{j}\nu _{b}) &\rightarrow &-i\mathrm{e}%
^{i(t_{0}-t_{b})\zeta _{j}^{-}},\qquad (\hat{\varkappa}\ell _{j}\hat{\nu}%
_{b})\rightarrow -i\mathrm{e}^{i(t_{0}-t_{b})\xi _{j}^{-}}, \\
(\nu _{a}\ell _{j}\varkappa ) &\rightarrow &i\mathrm{e}^{i(t_{a}-t_{0})\zeta
_{j}^{+}},\qquad \phantom{-}(\hat{\nu}_{a}\ell _{j}\varkappa )\rightarrow i%
\mathrm{e}^{i(t_{a}-t_{0})\xi _{j}^{+}}, \\
\quad (\nu _{a}\ell _{j}\hat{\varkappa}) &\rightarrow &i\mathrm{e}%
^{i(t_{a}-t_{0})\zeta _{j}^{+}},\qquad \phantom{-}(\hat{\nu}_{a}\ell _{j}%
\hat{\varkappa})\rightarrow i\mathrm{e}^{i(t_{a}-t_{0})\xi _{j}^{+}},
\end{eqnarray*}%
which shows that in all cases a marked $\varkappa $ gives the same
propagator as does an unmarked $\varkappa $. $\square $

\bigskip

The next step is to extract useful pieces of information from the result (%
\ref{zero}). The left-hand side of equation (\ref{theorem}) can be expanded
as a sum%
\begin{equation}
\sum_{\text{markings }M}(-1)^{m}\prod\limits_{i=1}^{I}p_{Mi}(E_{i})=\sum_{%
\theta }c_{\theta }\prod\nolimits_{i=1}^{I}\frac{1}{E_{i}-\theta _{i}}
\label{pollar}
\end{equation}%
of polar monomials%
\begin{equation}
\prod\nolimits_{i=1}^{I}\frac{1}{E_{i}-\theta _{i}},  \label{contra}
\end{equation}%
where $\theta $ is an assignment of polar numbers $\theta _{i}=\zeta
_{i}^{+} $, $\xi _{i}^{+}$, $\zeta _{i}^{-}$ or $\xi _{i}^{-}$ to the legs $%
\ell _{i}$ of the diagram and $c_{\theta }$ are numerical coefficients. When
we integrate on the loop energies,%
\begin{equation}
\sum_{\theta }c_{\theta }\int \prod\nolimits_{i=1}^{I}\frac{1}{E_{i}-\theta
_{i}}\prod\nolimits_{j=1}^{L}\frac{\mathrm{d}e_{j}}{2\pi },  \label{inte}
\end{equation}%
multiply by $(2\pi )\delta (\sum_{i=1}^{V}\mathcal{E}_{i})$ and take the
Fourier transform, we get the left-hand side of (\ref{zero}). If we make
these operations on a single polar monomial (\ref{contra}), we obtain a
contribution that is proportional to a product of $\theta $ functions times
various exponential factors. We schematically write it as 
\begin{equation}
\prod\nolimits_{j=1}^{I}\theta (\Delta t_{j})\mathrm{e}^{i\rho _{j}\Delta
t_{j}}=\Theta (t_{1},\ldots ,t_{I})\prod\nolimits_{j=1}^{I}\mathrm{e}^{i\rho
_{j}\Delta t_{j}},\qquad \text{where \quad }\Theta (t_{1},\ldots
,t_{I})\equiv \prod\nolimits_{j=1}^{I}\theta (\Delta t_{j}).  \label{contro}
\end{equation}%
Here $\rho _{j}$ can be $\zeta _{j}^{+}$, $\xi _{j}^{+}$, $-\zeta _{j}^{-}$
or $-\xi _{j}^{-}$, and $\Delta t_{j}=t_{a_{j}}-t_{b_{j}}$ in the first two
cases, $\Delta t_{j}=t_{b_{j}}-t_{a_{j}}$ in the other two. Note that each
product of exponential factors is associated with a unique distribution $%
\Theta (t_{1},\ldots ,t_{I})$.

We start from the knowledge that the left-hand side of (\ref{zero})
vanishes. We can isolate each contribution (\ref{contro}) from the others by
looking at the exponential factors. Since the numbers $\zeta _{j}^{+}$, $\xi
_{j}^{+}$, $\zeta _{j}^{-}$ and $\xi _{j}^{-}$ can be chosen arbitrarily,
apart from the signs of their imaginary parts, each contribution (\ref%
{contro}) must disappear independently from equation (\ref{zero}).

The Fourier transform (\ref{contro}) of a polar monomial can disappear from (%
\ref{zero}) for two reasons: the numerical coefficient $c_{\theta }$ in
front of it vanishes, or the distribution $\Theta (t_{1},\ldots ,t_{I})$ is
identically equal to zero. Consequently, the right-hand side of (\ref%
{theorem}) can only contain the polar monomials that have a vanishing $%
\Theta (t_{1},\ldots ,t_{I})$. Thus, it is mandatory to understand when that
happens.

\begin{figure}[t]
\begin{center}
\includegraphics[width=3.5truecm]{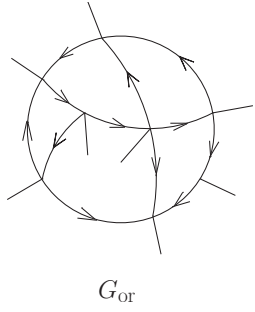}
\end{center}
\caption{Diagram with time ordered legs}
\label{gor}
\end{figure}

Consider the distribution $\Theta (t_{1},\ldots ,t_{I})$ together with the
\textquotedblleft naked\textquotedblright\ diagram $G$, that is to say the
diagram $G$ with no markings on the vertices and no orientations on the
lines. We want to use $\Theta (t_{1},\ldots ,t_{I})$ to equip $G$ with a 
\textit{time ordering} (which has nothing to do with the orientation based
on the energy flow met so far). Precisely, we equip each $G$ internal line
with an arrow pointing from the endpoint of lower time to the endpoint of
larger time. Denote the diagram obtained this way by $G_{\text{or}}$ (see
fig. \ref{gor}) and its distribution $\Theta (t_{1},\ldots ,t_{I})$ by $%
\Theta (G_{\text{or}})$.

We say that a curve $\gamma $ is time ordered if its lines have coherent
time orientations. We denote the product of the theta functions associated
with its lines by $\Theta (\gamma )$.

\begin{figure}[b]
\begin{center}
\includegraphics[width=9truecm]{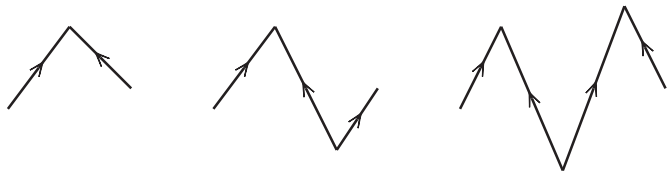}
\end{center}
\caption{Adjacent legs with opposite time orderings}
\label{timeor}
\end{figure}

If $\gamma $ is not time ordered and $t_{a}$, $t_{b}$ are the times
associated with its endpoints, $\Theta (\gamma )$ is nontrivial both for $%
t_{a}>t_{b}$ and $t_{a}<t_{b}$. To see this, observe that, since we are just
interested in the endpoints, two adjacent lines with coherent time orderings
can be collapsed onto a single line with the same ordering. Thus, it is
sufficient to consider the cases where adjacent legs have opposite time
orderings, as in the examples of fig. \ref{timeor}, where time is the
vertical coordinate. It is evident that for arbitrary $t_{a}$ and $t_{b}$,
there exist configurations of the intermediate vertices that make $\Theta
(\gamma )$ nontrivial.

Now we prove two useful lemmas. The first one is a generalization of the
property just shown.

\begin{lemma}
Let $\nu _{a}$ and $\nu _{b}$ denote two distinct vertices of $G_{\text{or}}$%
. Denote their times by $t_{a}$ and $t_{b}$, respectively. Assume that $%
\Theta (G_{\text{or}})$ is nontrivial, but vanishes identically for $%
t_{a}>t_{b}$. Then $G_{\text{or}}$ contains a time ordered curve that
connects $\nu _{a}$ to $\nu _{b}$. \label{lemma2}
\end{lemma}

\textbf{Proof}. The distribution $\Theta (G_{\text{or}})$ can be viewed as a
set constraints on the relative times of the nearest neighbors. When any of
those constraints is violated, $\Theta (G_{\text{or}})$ vanishes. We can
assume that $v_{a}$ and $v_{b}$ are not nearest neighbors, because in that
case the theorem is trivial. Assume that $\Theta (G_{\text{or}})$ forces a
vertex $\nu $, different from $\nu _{a}$ and $\nu _{b}$, to be in the future
of all its nearest neighbors. If so, send $\nu $ to the infinite future,
which is equivalent to dropping $\nu $ and cutting the legs attached to it.
Similarly, if $\Theta (G_{\text{or}})$ forces a vertex $\bar{\nu}\neq \nu
_{a},\nu _{b}$ to be in the past of all its nearest neighbors, send it to
the infinite past. Once both types of vertices are dropped, a reduced
diagram $G_{\text{or}}^{\prime }$ is obtained, equipped with a reduced
distribution $\Theta (G_{\text{or}}^{\prime })$. Since $\Theta (G_{\text{or}%
})$ vanishes identically for $t_{a}>t_{b}$, $\Theta (G_{\text{or}}^{\prime
}) $ satisfies the same property. Next, repeat the procedure on $G_{\text{or}%
}^{\prime }$: $\Theta (G_{\text{or}}^{\prime })$ may force other vertices,
different from $\nu _{a}$ and $\nu _{b}$, to be in the past or future of all
their nearest neighbors; if we drop them, we obtain a further reduced
diagram $G_{\text{or}}^{\prime \prime }$, on which we can iterate again. At
the end, we remain with a reduced diagram $G_{\text{or}}^{\text{red}}$ that
contains $\nu _{a}$, $\nu _{b}$ and possibly vertices that are forced to
have both past and future nearest neighbors. Moreover, $\Theta (G_{\text{or}%
}^{\text{red}})$ vanishes identically for $t_{a}>t_{b}$. In particular, $G_{%
\text{or}}^{\text{red}}$ cannot contain just $\nu _{a}$ and $\nu _{b}$,
because in that case the distribution $\Theta (G_{\text{or}}^{\text{red}})$
would be identically one. We infer that $G_{\text{or}}^{\text{red}}$ must
contain at least one vertex $\nu $ besides $\nu _{a}$ and $\nu _{b}$. Then, $%
\nu $ must have a future neighbor $\nu _{\text{f}}$ and a past neighbor $\nu
_{\text{p}}$. Similarly, $\nu _{\text{f}}$ must have a future neighbor $\nu
_{\text{f}}^{\prime }$, while $\nu _{\text{p}}$ must have a past neighbor $%
\nu _{\text{p}}^{\prime }$. Continuing like this, we eventually reach $\nu
_{b}$ in the future and $\nu _{a}$ in the past, and identify a time ordered
curve $\gamma _{ab}$ that connects $\nu _{a}$ to $\nu _{b}$. $\square $

\begin{lemma}
The distribution $\Theta (G_{\text{or}})$ is trivial if and only if $G_{%
\text{or}}$ contains a time ordered loop.
\end{lemma}

For example, the diagram of fig. \ref{gor} has a time ordered loop. A time
ordered loop clearly vanishes, because the theta functions conflict with one
another, as in%
\begin{equation}
\theta (t_{1}-t_{2})\theta (t_{2}-t_{1}),\qquad \theta (t_{1}-t_{2})\theta
(t_{2}-t_{3})\theta (t_{3}-t_{1}).  \label{orie}
\end{equation}%
The key content of the Lemma is that this is the only situation that can
make $\Theta (t_{1},\ldots ,t_{I})$ vanish.

\bigskip

\textbf{Proof}. Assume that $G_{\text{or}}$ is a tree diagram. Then, the
distribution $\Theta (G_{\text{or}})$ is nontrivial, because it just orders
the vertices according to time: the configurations $t_{1},\ldots ,t_{I}$
where $\Theta (t_{1},\ldots ,t_{I})$ is equal to one have nonvanishing
measure.

Now, proceed by induction. Assume that the theorem holds for diagrams with $%
L $ loops or less. Consider an $(L+1)$-loop diagram $G_{\text{or}}$. Cut one
leg $(\nu _{a}\bar{\ell}\nu _{b})$, so as to obtain an $L$ loop diagram $%
G_{L}$, which satisfies the theorem by the inductive assumption. If $\Theta
(G_{L})$ is trivial, it has a time oriented loop and so does $\Theta (G_{%
\text{or}})$. If $\Theta (G_{L})$ is nontrivial, we distinguish two cases: ($%
i$) $\Theta (G_{L})$ is nontrivial for both $t_{a}<t_{b}$ and $t_{a}>t_{b}$;
($ii$) $\Theta (G_{L})$ is trivial for either $t_{a}<t_{b}$ or $t_{a}>t_{b}$%
. When we close the $(L+1)$-th loop, the leg $(\nu _{a}\bar{\ell}\nu _{b})$
orders the times $t_{a}$ and $t_{b}$. In case ($i$), $\Theta (G_{\text{or}})$
is nontrivial. In case ($ii$), $\Theta (G_{\text{or}})$ is trivial if and
only if the time ordering due to $(\nu _{a}\bar{\ell}\nu _{b})$ conflicts
with the one due to $\Theta (G_{L})$. By Lemma \ref{lemma2}, $G_{L}$
contains a time ordered curve $\gamma _{ab}$ connecting $\nu _{a}$ and $\nu
_{b}$. Thus, $\Theta (G_{\text{or}})$ is trivial if and only if the union of 
$(\nu _{a}\bar{\ell}\nu _{b})$ and $\gamma _{ab}$ is a time oriented loop. $%
\square $

\bigskip

Now, let us go back to equation (\ref{pollar}), that is to say the expansion
of the left-hand side of (\ref{theorem}) in terms of polar monomials (\ref%
{contra}). We recall that every polar monomial leads to a contribution (\ref%
{contro}), when we integrate on the loop energies and take the Fourier
transform. Since each contribution (\ref{contro}) is independent of the
others, equation (\ref{zero}) implies that every time the distribution $%
\Theta (t_{1},\ldots ,t_{I})$ is nontrivial, the coefficient $c_{\theta }$
must vanish. Thus, the right-hand side of (\ref{pollar}) contains only the
polar monomials (\ref{contra}) that lead to a trivial $\Theta (t_{1},\ldots
,t_{I})$. Moreover, we have just proved that $\Theta (t_{1},\ldots ,t_{I})$
is trivial if and only if $G_{\text{or}}$ contains a time ordered loop $%
\gamma _{\text{or}}$, i.e. $\Theta (\gamma _{\text{or}})$ is trivial. Now we
have to understand how this requirement reflects on the polar monomial (\ref%
{contra}).

Consider the legs $\ell _{i}$ of $\gamma _{\text{or}}$ and collect the
values of their indices $i$ into the set $s_{\text{or}}$. Multiply $\Theta
(\gamma _{\text{or}})$ by the appropriate exponential factors $\mathrm{e}%
^{i\rho _{j}\Delta t_{j}}$, as in (\ref{contro}), 
\begin{equation}
\Theta (\gamma _{\text{or}})\prod\nolimits_{j\in s_{\text{or}}}\mathrm{e}%
^{i\rho _{j}\Delta t_{j}}=\prod\nolimits_{j\in s_{\text{or}}}\theta (\Delta
t_{j})\mathrm{e}^{i\rho _{j}\Delta t_{j}}.  \label{Tgor}
\end{equation}

We first assume that $\gamma _{\text{or}}$ is also oriented, in the sense of
the energy flow. Then, the triviality of the distribution (\ref{Tgor})
leads, after Fourier transform, to the identity%
\begin{equation}
\int_{-\infty }^{+\infty }\frac{\mathrm{d}e}{2\pi }\prod\nolimits_{i\in s_{%
\text{or}}}\frac{1}{E_{i}-\theta _{i}}=0,  \label{zeroc}
\end{equation}%
for arbitrary values of $\zeta _{i}^{+}$, $\zeta _{i}^{-}$, $\xi _{i}^{+}$
and $\xi _{i}^{-}$, $i\in s_{\text{or}}$, where $e$ is the loop energy of $%
\gamma _{\text{or}}$. Here, $e$ is one of the loop energies $e_{1},\ldots
,e_{L}$ and appears inside each $E_{i}$ of (\ref{zeroc}) with coefficient $%
+1 $.

Now we show that the identity (\ref{zeroc}) holds if and only if the polar
numbers $\theta _{i}$ are all placed on the same side with respect to the
real axis, which means that the loop $\gamma _{\text{or}}$ is polarized. It
is obvious that this condition is sufficient, because if we close the
integration path on the half plane with no poles, the residue theorem gives
zero. The condition is also necessary, as we show by reductio ad absurdum.
Assume that the integral of (\ref{zeroc}) is identically zero when one or
more poles are above the real axis and one or more poles are below it. Move
all the poles with positive imaginary parts into a single pole $\theta ^{+}$
and all those with negative imaginary parts into a single pole $\theta ^{-}$%
. This gives an integral of the form%
\begin{equation*}
\int_{-\infty }^{+\infty }\frac{1}{(e-\theta ^{+})^{n_{+}}(e-\theta
^{-})^{n_{-}}}\frac{\mathrm{d}e}{2\pi }=\binom{n_{+}+n_{-}-2}{n_{+}-1}\frac{%
i(-1)^{n_{+}+1}}{(\theta ^{+}-\theta ^{-})^{n_{+}+n_{-}-1}},
\end{equation*}%
which is obviously nonvanishing, contradicting the assumption. Thus, $\gamma
_{\text{or}}$ is polarized.

If $\gamma _{\text{or}}$ is not oriented in the sense of the energy flow,
its loop energy $e$ must be defined anew, since it is not one of the
standard integrations variables $e_{1},\ldots ,e_{L}$ we have been using so
far. Choose a direction for the $e$ flow along $\gamma _{\text{or}}$ and
split the set $s_{\text{or}}$ into $s_{\text{or}}^{\prime }\cup s_{\text{or}%
}^{\prime \prime }$ , such that the legs $\ell _{i}$ with $i\in s_{\text{or}%
}^{\prime }$ have orientations coherent with the $e$ flow, while the legs $%
\ell _{i}$ with $i\in $ $s_{\text{or}}^{\prime \prime }$ have orientations
opposite to the $e$ flow. Consider the integrand of (\ref{zeroc}) and write $%
E_{i}=e+E_{i}^{\prime }$ for $i\in s_{\text{or}}^{\prime }$, $%
E_{i}=-e+E_{i}^{\prime }$ for $i\in s_{\text{or}}^{\prime \prime }$, where $%
E_{i}^{\prime }$ are energies independent of $e$. Then, the condition that
the Fourier transform of (\ref{Tgor}) vanishes identically gives%
\begin{equation*}
\int_{-\infty }^{+\infty }\frac{\mathrm{d}e}{2\pi }\prod\nolimits_{i\in s_{%
\text{or}}^{\prime }}\frac{1}{E_{i}^{\prime }+e-\theta _{i}}%
\prod\nolimits_{j\in s_{\text{or}}^{\prime \prime }}\frac{1}{E_{j}^{\prime
}-e-\theta _{j}}=0.
\end{equation*}%
We know that this condition holds if and only if the poles are located on
the same side of the complex plane with respect to the real $e$ axis. This
means that each $\theta _{i}$ with $i\in s_{\text{or}}^{\prime }$ must be
located on one half plane and each $\theta _{i}$ with $i\in s_{\text{or}%
}^{\prime \prime }$ must be located on the other half plane. We see again
that the loop $\gamma _{\text{or}}$ is polarized, i.e. adjacent $\gamma _{%
\text{or}}$ legs of coherent (opposite) orientations carry polar numbers of
coherent (opposite) polarities.

Since the conclusions hold for arbitrary energies, as well as arbitrary
polar numbers $\zeta _{i}^{+}$, $\xi _{i}^{+}$, $\zeta _{i}^{-}$ and $\xi
_{i}^{-}$, it also holds for arbitrary polar numbers $\sigma _{i}^{+}$, $%
\tau _{i}^{+}$, $\sigma _{i}^{-}$ and $\tau _{i}^{-}$. This gives formula (%
\ref{theorem}) and concludes the proof. $\square $

\section{Examples and applications}

\setcounter{equation}{0}\label{examples}

In quantum field theory, we can decompose each propagator into the sum of
two polar numbers, called \textquotedblleft half
propagators\textquotedblright , each of which has a unique pole. The
polarity refers to the location of the pole with respect to the real axis.
From now on, positive (negative) polarity means that the pole is located
below (above) the real axis.

A polarized monomial has a polarized loop. As explained above, the integral
on the energy of a polarized loop is equal to zero, because all the poles of
its integrand are located on the same side with respect to the real axis.
Therefore, the integral of the left-hand side of (\ref{theorem}) on the loop
momenta vanishes. This operation leads to the diagrammatic cutting equation
associated with the diagram $G$.

In this section, we illustrate these properties in various one-loop and
two-loop diagrams and include the algebraic identities of other diagrams, up
to three loops. In section \ref{s5} we generalize them to prove the
perturbative unitarity of quantum field theories.

\subsection{Bubble diagram}

The \textquotedblleft bubble\textquotedblright\ diagram is the diagram ($b$)
of fig. \ref{orientation}. Its marked versions are shown in fig. \ref{polar}
and lead to the polynomial identity (\ref{deco}). Now we show how to apply
this identity and derive the diagrammatic cutting equations.

The value of the bubble diagram is given by the convolution of two
propagators. In $D$ dimensional scalar field theories, we have%
\begin{equation}
\mathcal{B}=\int \frac{\mathrm{d}^{D}k}{(2\pi )^{D}}\frac{1}{%
k^{2}-m_{1}^{2}+i\epsilon }\frac{1}{(k-p)^{2}-m_{2}^{2}+i\epsilon ^{\prime }}%
.  \label{bubba}
\end{equation}%
For convenience, we keep the infinitesimal widths $\epsilon $ and $\epsilon
^{\prime }$ different from each other. The reason will become apparent
below. The arguments that follow focus on the energy integral, which is
convergent. We do not need to pay attention to the integral on the space
momentum. That integral may diverge in certain dimensions $D$, in which case
it can be defined by means of a regularization (the dimensional technique
being the most convenient choice).

Define the polar numbers%
\begin{equation}
\sigma _{1}^{\pm }=\pm \frac{1}{2\omega _{1\epsilon }}\frac{1}{k^{0}\mp
\omega _{1\epsilon }},\qquad \sigma _{2}^{\pm }=\pm \frac{1}{2\omega
_{2\epsilon ^{\prime }}}\frac{1}{k^{0}-p^{0}\mp \omega _{2\epsilon ^{\prime
}}},\qquad \tau _{i}^{\pm }=-(\sigma _{i}^{\mp })^{\ast },\qquad
\label{pol1}
\end{equation}%
where the complex frequencies are $\omega _{1\epsilon }=\sqrt{\mathbf{k}%
^{2}+m_{1}^{2}-i\epsilon }$, $\omega _{2\epsilon ^{\prime }}=\sqrt{(\mathbf{%
k-p)}^{2}+m_{2}^{2}-i\epsilon ^{\prime }}$ and contain the $\epsilon
,\epsilon ^{\prime }$ prescriptions. We see that $\sigma _{i}^{+}$ and $\tau
_{i}^{+}$ have poles located below the real axis, while $\sigma _{i}^{-}$
and $\tau _{i}^{-}$ have poles located above the real axis.

Note that the definition of polarity we use here differs from the one used
in the proof of the previous section in several respects. In particular, the
signs of the imaginary parts of $\sigma _{i}^{\pm }$ and $\tau _{i}^{\pm }$
do not agree with the signs of the imaginary parts of their poles. We recall
that the algebraic theorem of section \ref{s2} works with any definition of
polarity.

The combinations%
\begin{equation}
z_{1}=\sigma _{1}^{+}+\sigma _{1}^{-}=\frac{1}{k^{2}-m_{1}^{2}+i\epsilon }%
,\qquad z_{2}=\sigma _{2}^{+}+\sigma _{2}^{-}=\frac{1}{(k-p)^{2}-m_{2}^{2}+i%
\epsilon ^{\prime }},\qquad w_{i}=-z_{i}^{\ast },  \label{pro1}
\end{equation}%
give the propagators and (minus) their conjugates.

Now, define the \textquotedblleft cut propagators\textquotedblright\ $u_{i}$
and $v_{i}$ as 
\begin{equation}
u_{i}=\sigma _{i}^{+}+\tau _{i}^{-},\qquad v_{i}=\sigma _{i}^{-}+\tau
_{i}^{+}.  \label{pra1}
\end{equation}%
We call these combinations cut propagators\ even if they are defined at $%
\epsilon ,\epsilon ^{\prime }\neq 0$. Strictly speaking, the usual cut
propagators are obtained in the limits $\epsilon ,\epsilon ^{\prime
}\rightarrow 0$. For example, using $\omega _{1\epsilon }\sim \omega
_{1}-i\epsilon /(2\omega _{1})$, where $\omega _{1}=\omega _{1\epsilon
}|_{\epsilon =0}$, we get 
\begin{equation}
\lim_{\epsilon \rightarrow 0}u_{1}=-2i\pi \theta (k^{0})\delta
(k^{2}-m_{1}^{2}),\qquad \lim_{\epsilon \rightarrow 0}v_{1}=-2i\pi \theta
(-k^{0})\delta (k^{2}-m_{1}^{2}),  \label{cutpr}
\end{equation}%
which are the usual cut propagators of a scalar field, multiplied by $-i$.
The limits of $u_{2}$ and $v_{2}$ for $\epsilon ^{\prime }\rightarrow 0$
give (\ref{cutpr}) with the replacements $k\rightarrow k-p$ and $%
m_{1}\rightarrow m_{2}$.

\bigskip

\begin{figure}[t]
\begin{center}
\includegraphics[width=14truecm]{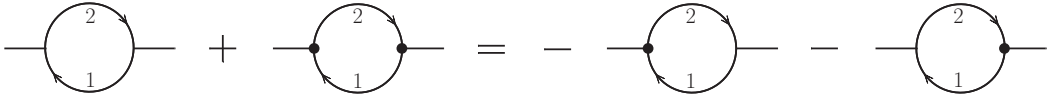}
\end{center}
\caption{Cutting equation of the bubble diagram}
\label{bubb}
\end{figure}

At this point, we can write the bubble diagram (\ref{bubba}) and its
conjugate in the form%
\begin{equation*}
\mathcal{B}=\int z_{1}z_{2},\qquad \mathcal{B}^{\ast }=\int w_{1}w_{2},
\end{equation*}%
and use the polynomial identity (\ref{deco}). The decomposition (\ref{deco})
is advantageous for the integration over the loop energy $k^{0}$. Expand the
right-hand side of (\ref{deco}) as a sum of polarized monomials and pick one
such monomial at a time. Its poles are located on the same side of the
complex plane with respect to the real axis. When we integrate $k^{0}$ along
the real axis, we can close the integration path at infinity on the side
that contains no poles. By the residue theorem, each polarized monomial
gives zero. Thus, the momentum integral of the left-hand side of (\ref{deco}%
) also vanishes. This gives the relation%
\begin{equation}
\mathcal{B}+\mathcal{B}^{\ast }=\int u_{1}v_{2}+\int v_{1}u_{2},
\label{bubbi}
\end{equation}%
which is graphically shown in fig. \ref{bubb}. The Feynman rules are those
of fig. \ref{feybub} (together with a factor $-1$ for every marked vertex).

\begin{figure}[b]
\begin{center}
\includegraphics[width=8truecm]{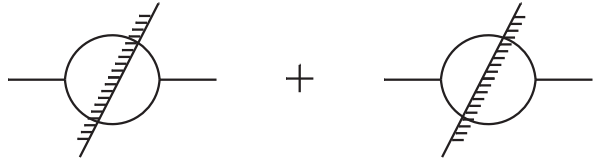}
\end{center}
\caption{Cut bubble diagrams}
\label{grafici}
\end{figure}

The right-hand side of fig. \ref{bubb} is minus the sum of the cut diagrams,
which is popularly represented as shown in fig. \ref{grafici} by shadowing
the areas that contain the marked vertices.

Note that $\epsilon ,\epsilon ^{\prime }$ are still different from zero, so
the identity (\ref{bubbi}) is actually more general than the ones we are
accustomed to in quantum field theory. Indeed, in those identities, the cut
propagators $u_{i}$ are replaced by their limits (\ref{cutpr}). In our
identity, instead, $\epsilon $ and $\epsilon ^{\prime }$ can be arbitrary
positive numbers.

We see that the diagrammatic cutting equation of fig. \ref{bubb} is a
straightforward consequence of the simple polynomial identity (\ref{deco}).

\subsection{Triangle and box diagrams}

\begin{figure}[t]
\begin{center}
\includegraphics[width=14truecm]{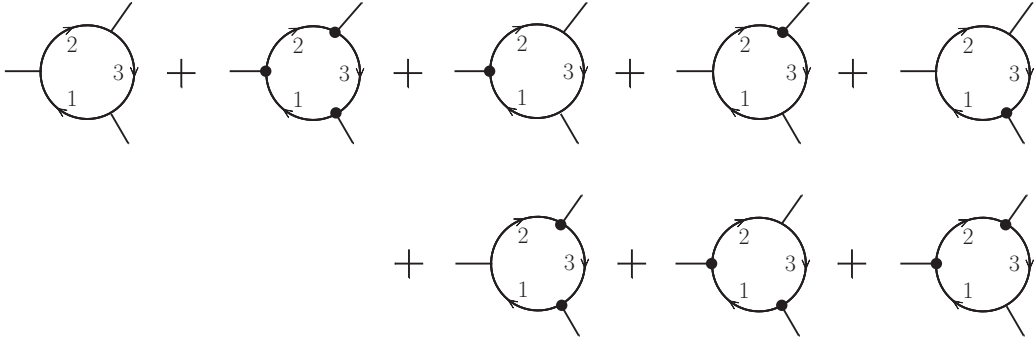}
\end{center}
\caption{Marked triangle diagrams}
\label{trag}
\end{figure}

The \textquotedblleft triangle\textquotedblright\ diagram is the one-loop
oriented diagram with three adjacent lines. It leads to the polynomial
identity%
\begin{eqnarray}
&&z_{1}z_{2}z_{3}-w_{1}w_{2}w_{3}-u_{1}v_{2}z_{3}-z_{1}u_{2}v_{3}-v_{1}z_{2}u_{3}+v_{1}u_{2}w_{3}+w_{1}v_{2}u_{3}+u_{1}w_{2}v_{3}
\notag \\
&&\qquad \qquad =\prod\nolimits_{i=1}^{3}(\sigma _{i}^{+}-\tau
_{i}^{+})+\prod\nolimits_{i=1}^{3}(\sigma _{i}^{-}-\tau _{i}^{-}),
\label{polo}
\end{eqnarray}%
which is easy to verify directly. The left-hand side of this formula can be
worked out from the sum of the triangle diagram plus its marked versions,
shown in fig. \ref{trag}, by applying the Feynman rules of fig. \ref{feybub}
and multiplying by $(-1)^{m}$, where $m$ is the number of marked vertices.
The right-hand side of (\ref{polo}) corresponds of the polynomial $\mathcal{P%
}_{G}$ of formula (\ref{theorem}), which in the one-loop case is known in
closed form due to formula (\ref{wou}).

The triangle diagram gives the integral%
\begin{equation}
\mathcal{T}=\int \frac{\mathrm{d}^{D}k}{(2\pi )^{D}}\frac{1}{%
k^{2}-m_{1}^{2}+i\epsilon }\frac{1}{(k-p)^{2}-m_{2}^{2}+i\epsilon ^{\prime }}%
\frac{1}{(k-q)^{2}-m_{3}^{2}+i\epsilon ^{\prime \prime }},  \label{triangle}
\end{equation}%
where $p$ and $q$ are external momenta. To extend the analysis of the
previous section, we add the definitions%
\begin{eqnarray}
\sigma _{3}^{\pm } &=&\pm \frac{1}{2\omega _{3\epsilon ^{\prime \prime }}}%
\frac{1}{k^{0}-q^{0}\mp \omega _{3\epsilon ^{\prime \prime }}},\qquad
z_{3}=\sigma _{3}^{+}+\sigma _{3}^{-},\qquad \tau _{3}^{\pm }=-(\sigma
_{3}^{\mp })^{\ast },\qquad w_{3}=-z_{3}^{\ast },  \notag \\
u_{3} &=&\sigma _{3}^{+}+\tau _{3}^{-},\qquad v_{3}=\sigma _{3}^{-}+\tau
_{3}^{+},  \label{pol2}
\end{eqnarray}%
to the previous ones, where $\omega _{3\epsilon ^{\prime \prime }}=\sqrt{(%
\mathbf{k-q)}^{2}+m_{3}^{2}-i\epsilon ^{\prime \prime }}$.

Again, we integrate both members of equation (\ref{polo}) on $k^{0}$. The
right-hand side gives zero by the residue theorem, because of our definition
of polarity, while the left-hand side leads to the identity%
\begin{equation*}
\mathcal{T}+\mathcal{T}^{\ast }-\int u_{1}v_{2}z_{3}-\int
z_{1}u_{2}v_{3}-\int v_{1}z_{2}u_{3}+\int v_{1}u_{2}w_{3}+\int
w_{1}v_{2}u_{3}+\int u_{1}w_{2}v_{3}=0.
\end{equation*}%
This formula is graphically represented by equating the sum of fig. \ref%
{trag} to zero. It can also be viewed as a diagrammatic cutting equation, by
shadowing the marked areas of the diagrams. Once again, the diagrammatic
cutting equation is rooted into the simple polynomial identity (\ref{polo}).

Without giving further details, we report the polynomial identity associated
with the box diagram, which is%
\begin{eqnarray*}
&&z_{1}z_{2}z_{3}z_{4}+w_{1}w_{2}w_{3}w_{4}-u_{1}v_{2}z_{3}z_{4}+u_{1}w_{2}v_{3}z_{4}-u_{1}w_{2}w_{3}v_{4}-z_{1}u_{2}v_{3}z_{4}+z_{1}u_{2}w_{3}v_{4}-v_{1}u_{2}w_{3}w_{4}
\\
&&-z_{1}z_{2}u_{3}v_{4}+v_{1}z_{2}u_{3}w_{4}-w_{1}v_{2}u_{3}w_{4}-v_{1}z_{2}z_{3}u_{4}+w_{1}v_{2}z_{3}u_{4}-w_{1}w_{2}v_{3}u_{4}+v_{1}u_{2}v_{3}u_{4}+u_{1}v_{2}u_{3}v_{4}
\\
&&\qquad =\prod\nolimits_{i=1}^{4}(\sigma _{i}^{+}-\tau
_{i}^{+})+\prod\nolimits_{i=1}^{4}(\sigma _{i}^{-}-\tau _{i}^{-}).
\end{eqnarray*}

\subsection{Two-loop and three-loop diagrams}

\begin{figure}[t]
\begin{center}
\includegraphics[width=14truecm]{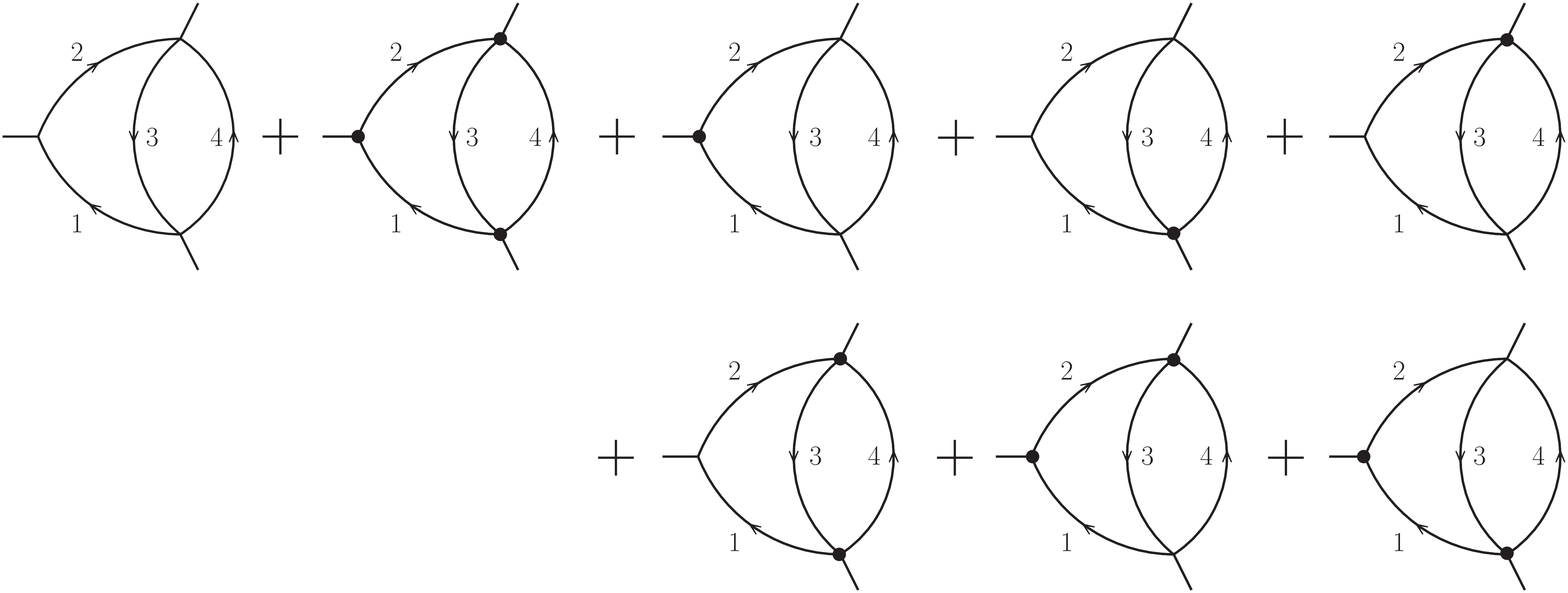}
\end{center}
\caption{Marked chestnut diagrams}
\label{chestnut}
\end{figure}

Now we give a two-loop example, the chestnut diagram shown as first in fig. %
\ref{chestnut}, which includes its marked versions. The associated
polynomial identity reads%
\begin{eqnarray}
&&z_{1}z_{2}z_{3}z_{4}-w_{1}w_{2}w_{3}w_{4}-u_{1}v_{2}z_{3}z_{4}-v_{1}z_{2}u_{3}v_{4}-z_{1}u_{2}v_{3}u_{4}+v_{1}u_{2}w_{3}w_{4}+u_{1}w_{2}v_{3}u_{4}+w_{1}v_{2}u_{3}v_{4}
\notag \\
&&\qquad \qquad \qquad \qquad \qquad =\sum \left[ a\rho _{1}^{+}\rho
_{2}^{+}\rho _{3}^{+}\eta _{4}+b\rho _{1}^{-}\rho _{2}^{-}\rho _{3}^{-}\eta
_{4}+c\eta _{1}\eta _{2}\rho _{3}^{+}\rho _{4}^{+}+d\eta _{1}\eta _{2}\rho
_{3}^{-}\rho _{4}^{-}\right] ,  \label{chest}
\end{eqnarray}%
where each $\rho _{i}$ can stand for $\sigma _{i}$ or $\tau _{i}$ and each $%
\eta _{i}$ can stand for $\sigma _{i}^{+}$, $\sigma _{i}^{-}$, $\tau
_{i}^{+} $ or $\tau _{i}^{-}$. The sum is over all such choices, $a,b,c,d$
denoting unspecified numerical coefficients.

The oriented loops are 123 and 34. The loop 124 is redundant, because
whenever it is polarized, either 123 or 34 is also polarized. The right-hand
side of (\ref{chest}) is a sum of polarized monomials that factorize the
polarized loops $\rho _{1}^{+}\rho _{2}^{+}\rho _{3}^{+}$, $\rho
_{1}^{-}\rho _{2}^{-}\rho _{3}^{-}$, $\rho _{3}^{+}\rho _{4}^{+}$ or $\rho
_{3}^{-}\rho _{4}^{-}$, as required by formula (\ref{theorem}).

The chestnut diagram gives the loop integral%
\begin{equation*}
\mathcal{C}=\int \mathrm{d}\mu \frac{1}{k^{2}-m_{1}^{2}+i\epsilon _{1}}\frac{%
1}{(k-p)^{2}-m_{2}^{2}+i\epsilon _{2}}\frac{1}{(k+q-p^{\prime
})^{2}-m_{3}^{2}+i\epsilon _{3}}\frac{1}{q^{2}-m_{4}^{2}+i\epsilon _{4}},
\end{equation*}%
where $p$ and $p^{\prime }$ are external momenta and the measure $\mathrm{d}%
\mu $ is $\mathrm{d}^{D}k\mathrm{d}^{D}q/(2\pi )^{2D}$. The definitions of
polar numbers, propagators and cut propagators are straightforward,
mimicking the formulas (\ref{pol1}), (\ref{pro1}), (\ref{pra1}) and (\ref%
{pol2}). When we integrate on the loop momenta, the right-hand side of (\ref%
{chest}) gives zero, since every term contains a polarized oriented loop.
The integral on the energy of that loop vanishes by the residue theorem,
since the integrand has poles only above or below the real axis. In the end,
we obtain the diagrammatic cutting equation graphically represented by
equating the sum of fig. \ref{chestnut} to zero.

If we flip the orientations of the legs 3 and 4, we obtain a different
orientation, for the diagram, and a different polynomial identity, which is
equal to (\ref{chest}) upon exchange of the subscripts 3 and 4. In that
case, the oriented loops become\ 124 and 34, so the right-hand side of (\ref%
{chest}) contains the polarized factors $\rho _{1}^{+}\rho _{2}^{+}\rho
_{4}^{+}$, $\rho _{1}^{-}\rho _{2}^{-}\rho _{4}^{-}$, $\rho _{3}^{+}\rho
_{4}^{+}$ and $\rho _{3}^{-}\rho _{4}^{-}$. The two orientations lead to
equivalent identities for the integral $\mathcal{C}$, because they amount to
send the loop momentum $q$ to $-k-q$. To better see this, it is convenient
to switch off the external momenta $p$ and $p^{\prime }$, because they are
not important for the polynomial identity.

We also report the polynomial identities associated with the two-loop
self-energy diagrams of fig. \ref{polloop}. The first diagram gives%
\begin{eqnarray}
&&z_{1}z_{2}z_{3}z_{4}z_{5}+w_{1}w_{2}w_{3}w_{4}w_{5}-u_{1}v_{2}z_{3}z_{4}z_{5}-z_{1}u_{2}v_{3}z_{4}u_{5}-z_{1}z_{2}z_{3}u_{4}v_{5}-v_{1}z_{2}u_{3}v_{4}z_{5}
\notag \\
&&+u_{1}w_{2}v_{3}z_{4}u_{5}+u_{1}v_{2}z_{3}u_{4}v_{5}+w_{1}v_{2}u_{3}v_{4}z_{5}+v_{1}u_{2}w_{3}v_{4}u_{5}+z_{1}u_{2}v_{3}u_{4}w_{5}+v_{1}z_{2}u_{3}w_{4}v_{5}
\notag \\
&&-v_{1}u_{2}w_{3}w_{4}w_{5}-w_{1}v_{2}u_{3}w_{4}v_{5}-w_{1}w_{2}w_{3}v_{4}u_{5}-u_{1}w_{2}v_{3}u_{4}w_{5}
\notag \\
&&\qquad =\sum \left[ a\rho _{1}^{+}\rho _{2}^{+}\rho _{3}^{+}\eta _{4}\eta
_{5}+b\rho _{1}^{-}\rho _{2}^{-}\rho _{3}^{-}\eta _{4}\eta _{5}+c\eta
_{1}\eta _{2}\rho _{3}^{+}\rho _{4}^{+}\rho _{5}^{+}+d\eta _{1}\eta _{2}\rho
_{3}^{-}\rho _{4}^{-}\rho _{5}^{-}\right] .  \label{last}
\end{eqnarray}%
The second diagram gives the same identity with $\sigma
_{4}^{+}\leftrightarrow \sigma _{4}^{-}$, $\tau _{4}^{+}\leftrightarrow \tau
_{4}^{-}$, $\sigma _{5}^{+}\leftrightarrow \sigma _{5}^{-}$, $\tau
_{5}^{+}\leftrightarrow \tau _{5}^{-}$.\ In either case, the loop 1245 is
redundant, because when it is polarized, either 123 or 345 is polarized.
Note that in the second diagram the loop 345 is not oriented and the last
two contributions of (\ref{last}) become proportional to the polarized loops 
$\rho _{3}^{+}\rho _{4}^{-}\rho _{5}^{-}$ and $\rho _{3}^{-}\rho
_{4}^{+}\rho _{5}^{+}$.

Finally, we give a three-loop example, the box diagram equipped with
diagonals. Let 1, 2, 3, 4 label the legs of the box and 5, 6 the diagonals.
Define the leg orientations so that the oriented loops are 1234, 125 and
236. Then the identity%
\begin{eqnarray*}
&&z_{1}z_{2}z_{3}z_{4}z_{5}z_{6}-u_{1}v_{2}z_{3}z_{4}z_{5}u_{6}+u_{1}w_{2}v_{3}z_{4}v_{5}u_{6}-u_{1}w_{2}w_{3}v_{4}v_{5}w_{6}-z_{1}u_{2}v_{3}z_{4}v_{5}z_{6}+z_{1}u_{2}w_{3}v_{4}v_{5}v_{6}
\\
&&-v_{1}u_{2}w_{3}w_{4}w_{5}v_{6}-z_{1}z_{2}u_{3}v_{4}z_{5}v_{6}+v_{1}z_{2}u_{3}w_{4}u_{5}v_{6}-w_{1}v_{2}u_{3}w_{4}u_{5}w_{6}-v_{1}z_{2}z_{3}u_{4}u_{5}z_{6}
\\
&&+w_{1}v_{2}z_{3}u_{4}u_{5}u_{6}-w_{1}w_{2}v_{3}u_{4}w_{5}u_{6}+v_{1}u_{2}v_{3}u_{4}w_{5}z_{6}+u_{1}v_{2}u_{3}v_{4}z_{5}w_{6}+w_{1}w_{2}w_{3}w_{4}w_{5}w_{6}\sim 0
\end{eqnarray*}%
holds, where the right-hand side of (\ref{theorem}), which we do not report
in full form, is a sum of polarized monomials. The polarized loops are 125,
236, 345 and 146, the last two being nonoriented.

\section{Perturbative unitarity of quantum field theories}

\setcounter{equation}{0}\label{s5}

In this section we show how to use the algebraic cutting equations to prove
the perturbative unitarity of quantum field theories. We begin with
nonderivative scalar theories. For definiteness, we may consider the $%
\varphi ^{4}$ theory, described by the Lagrangian%
\begin{equation*}
\mathcal{L}=\frac{1}{2}(\partial _{\mu }\varphi )(\partial ^{\mu }\varphi )-%
\frac{m^{2}}{2}\varphi ^{2}-\frac{\lambda }{4!}\varphi ^{4},
\end{equation*}%
which is renormalizable in $D\leqslant 4$. Alternatively, we may take the $%
\varphi ^{6}$ theory, which is renormalizable in $D\leqslant 3$, or the $%
\varphi ^{3}$ theory, which is renormalizable in $D\leqslant 6$. The
discussion is actually independent of the form of the potential and the
number of legs carried by the vertices, as long as they do not contain
derivatives. Derivative vertices may be included with a few extra
manipulations, which we describe at the end of this section. There, we also
generalize the arguments to fields of different spins and nonrenormalizable
theories.

Define the polar numbers%
\begin{equation}
\sigma _{\mathbf{k}\epsilon }^{\pm }=\pm \frac{1}{2\omega _{\mathbf{k}%
\epsilon }}\frac{1}{k^{0}\mp \omega _{\mathbf{k}\epsilon }},\qquad \tau _{%
\mathbf{k}\epsilon }^{\pm }=-(\sigma _{\mathbf{k}\epsilon }^{\mp })^{\ast
},\qquad  \label{compl}
\end{equation}%
where $\omega _{\mathbf{k}\epsilon }=\sqrt{\mathbf{k}^{2}+m^{2}-i\epsilon }$%
. The poles of $\sigma _{\mathbf{k}\epsilon }^{+}$, $\tau _{\mathbf{k}%
\epsilon }^{+}$ are located below the real axis and those of $\sigma _{%
\mathbf{k}\epsilon }^{-}$, $\tau _{\mathbf{k}\epsilon }^{-}$ are located
above the real axis. The combinations%
\begin{equation*}
z_{\mathbf{k}\epsilon }=\sigma _{\mathbf{k}\epsilon }^{+}+\sigma _{\mathbf{k}%
\epsilon }^{-}=\frac{1}{k^{2}-m^{2}+i\epsilon },\qquad w_{\mathbf{k}\epsilon
}=-z_{\mathbf{k}\epsilon }^{\ast },\qquad u_{\mathbf{k}\epsilon }=\sigma _{%
\mathbf{k}\epsilon }^{+}-(\sigma _{\mathbf{k}\epsilon }^{+})^{\ast },\qquad
v_{\mathbf{k}\epsilon }=\sigma _{\mathbf{k}\epsilon }^{-}-(\sigma _{\mathbf{k%
}\epsilon }^{-})^{\ast },
\end{equation*}%
give the propagators, their conjugates and the cut propagators. As before,
we can use a different $\epsilon $ for each propagator.

Given a Feynman diagram $G$ with $V$ vertices and $I$ internal legs, we
assign loop energies and an orientation to it as specified by proposition %
\ref{p1} of section \ref{s1}. We can promote the energy assignments to
assignments for the full momenta of the internal legs. So doing, we obtain a
parametrization of $G$ in momentum space.

\begin{figure}[t]
\begin{center}
\includegraphics[width=10truecm]{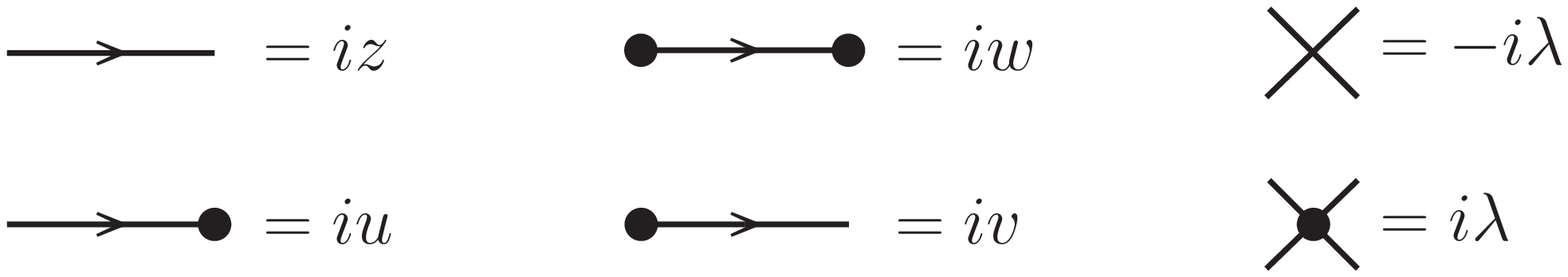}
\end{center}
\caption{Feynman rules in the standard notation}
\label{feyred}
\end{figure}

The Feynman rules we have used so far are made of the propagators of fig. %
\ref{feybub} plus the rule that an unmarked vertex is equal to $+1$ and a
marked vertex is equal to $-1$. The Feynman rules commonly used in quantum
field theory are slightly different, since they have an extra factor $i$ for
each propagator (cut or not) and an extra factor $-i\lambda $ for each
vertex (marked or not), $\lambda $ being the coupling. We show them in fig. %
\ref{feyred}. Each time we compute a diagram with the rules of the previous
sections, we miss the overall factor $i^{I}(-i\lambda )^{V}$ with respect to
the more common notation of fig. \ref{feyred}. For the rest of this section,
we switch to the common notation.

After these redefinitions, the propagator $iw$ connecting two marked points
is the complex conjugate of the propagator $iz$ connecting two unmarked
points. Moreover, the propagators $iu$ and $iv$ connecting a marked point to
an unmarked one are real. Finally, the marked vertices are the complex
conjugates of the unmarked vertices.

Now we turn to the identity (\ref{theorem}) and integrate it on the loop
momenta. If $G$ contains no tadpoles, the right-hand side vanishes, because
it is a sum of polarized monomials. We recall that a polarized monomial has
a polarized loop $\gamma_{\mathrm{pol}}$. As explained in section \ref{s3},
we can reparametrize the loop integral so that the energy $e$ of $\gamma_{%
\mathrm{pol}}$ is one of the integrated variables. The integral on $e$ is
zero by the residue theorem, since its integrand has poles only above or
below the real axis and we can close the integration path on the half plane
that contains no poles. The integral of the left-hand side of (\ref{theorem}%
) thus also vanishes, which gives the cutting equation. Once we multiply the
identity by the factors $i^{I+V}(-\lambda )^{V}$ and switch to the notation
of fig. \ref{feyred}, we arrive at the common diagrammatic cutting equation%
\begin{equation}
G(p_{1},\cdots ,p_{n})+\bar{G}(p_{1},\cdots ,p_{n})=-\sum_{\text{proper
markings }M}G_{M}(p_{1},\cdots ,p_{n}),  \label{mar}
\end{equation}%
where $p_{1},\cdots ,p_{n}$ are the external momenta, $G$ is the diagram
with all unmarked vertices, $\bar{G}$ is the diagram with all marked
vertices, and $G_{M}$ denotes a diagram with a \textquotedblleft
proper\textquotedblright\ marking, i.e. with at least one marked vertex and
one unmarked vertex.

Strictly speaking, (\ref{mar}) is the common cutting equation only in the
limit $\epsilon \rightarrow 0$, where the cut propagators force the energy
to flow from the unmarked endpoints to the marked endpoints. In that limit,
numerous marked diagrams vanish due to energy conservation. Those which
survive are precisely the usual cut diagrams. Nevertheless, it is always
possible to view a marked diagram as a cut diagram by means of closed cuts
that circle subdiagrams made of marked vertices. This way, we can extend the
common terminology by calling equation (\ref{mar}) a cutting equation even
at $\epsilon \neq 0$.

There is a caveat, though: we know that the diagrams that contain tadpoles
are not covered by the theorem of section \ref{s2}. Thus, as far as we know
now, equation (\ref{mar}) only holds for diagrams that contain no tadpoles.
We can extend formula (\ref{mar}) to the whole set of diagrams as follows.

\begin{figure}[t]
\begin{center}
\includegraphics[width=6truecm]{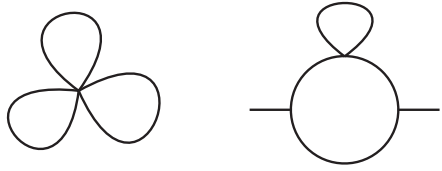}
\end{center}
\caption{Diagrams with tadpoles}
\label{tadpole}
\end{figure}

Tadpoles are loop diagrams with a unique vertex, $V=1$, so they have as many
internal lines as loops, by the topological identity $L-I+V=1$. A three-loop
example of a tadpole and a two-loop example of diagram with a tadpole are
shown in fig. \ref{tadpole}. We recall that the reason why the diagrams with
tadpoles are not covered by the theorem is that tadpoles lead to the
integrals of single polar numbers, which are not convergent. Indeed, they
behave as $\int \mathrm{d}k^{0}/k^{0}$ for $k^{0}$ large.

\begin{figure}[b]
\begin{center}
\includegraphics[width=14truecm]{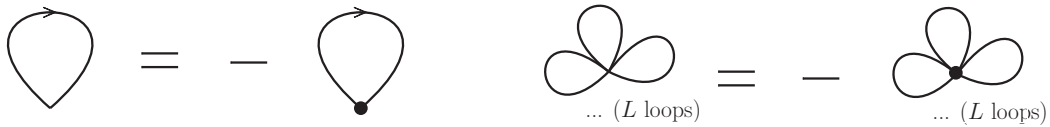}
\end{center}
\caption{Tadpole assumption}
\label{tad}
\end{figure}

Tadpoles and diagrams with tadpoles can be straightforwardly included in the
treatment, as long as they satisfy one additional assumption, which we call
the \textit{tadpole assumption}: the value of a tadpole with an unmarked
vertex must be opposite to the value of its marked version. Graphically, we
have fig. \ref{tad}.

\begin{figure}[t]
\begin{center}
\includegraphics[width=7truecm]{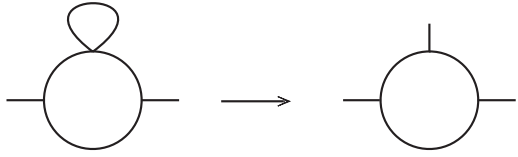}
\end{center}
\caption{Tadpole assumption}
\label{tadd}
\end{figure}

Now we show that the tadpole assumption allows us to derive the cutting
equations satisfied by the diagrams that contain tadpoles. Consider such a
diagram, and call it $G_{T}$. If we replace its tadpoles with external legs,
we obtain a diagram $G_{\hat{T}}$ that satisfies (\ref{mar}).

For example, let $G_{T}$ denote the second diagram of fig. \ref{tadpole}. If
we replace the tadpole subdiagram with an external leg, we obtain a diagram $%
G_{\hat{T}}$ that is equivalent to the triangle diagram treated in the
previous section (see fig. \ref{tadd}). We know that $G_{\hat{T}}$ satisfies
the identity obtained by equating the sum of fig. \ref{trag} to zero, which
leads to formula (\ref{mar}). Now, take fig. \ref{trag} and consider the
upper-right external leg and the vertex $\nu $ to which it is attached.
Suppress that leg and glue the tadpole to the vertex $\nu $. Thanks to the
first identity of fig. \ref{tad}, it does not matter whether $\nu $ is
marked or not, since the value of the tadpole (neglecting the minus sign due
to the marked vertex, because it is already counted inside $G_{\hat{T}}$) in
the same in both cases. This means that the tadpole attachment amounts to
multiplying the sum of fig. \ref{trag} by an overall factor. What we obtain
by doing this and equating the total to zero is precisely the cutting
equation satisfied by $G_{T}$.

The property illustrated in this simple example can be generalized to all
tadpole diagrams $G_{T}$, as long as the tadpole assumption holds.
Ultimately, to derive the cutting equations of tadpole diagrams, it is
sufficient to\ ignore the tadpole subdiagrams and apply the procedure used
for every other diagram.

It is easy to check that the scalar theories we are considering satisfy the
tadpole assumption, if the dimensional regularization is used and $\epsilon $
is sent to zero. For example, the value of the one-loop tadpole is [with the
rules of fig. \ref{feyred}] 
\begin{equation}
\int \frac{\mathrm{d}^{D}k}{(2\pi )^{D}}\frac{i(-i\lambda )}{%
k^{2}-m^{2}+i\epsilon }=-i\lambda \frac{\Gamma \left( 1-\frac{D}{2}\right) }{%
(4\pi )^{D/2}}(m^{2}-i\epsilon )^{(D-2)/2}\rightarrow -i\lambda \frac{\Gamma
\left( 1-\frac{D}{2}\right) }{(4\pi )^{D/2}}m^{D-2},  \label{umma}
\end{equation}%
while its marked version has the opposite value. The reason is that the
marked version carries a minus sign due to the marked vertex and two other
minus signs that compensate each other. Recall that the propagator
connecting two marked points is $iw=-iz^{\ast }$. The conjugation of $z$
flips the sign of the prescription $+i\epsilon $, which leads to a factor $%
-1 $ that compensates the minus sign in front of $iz^{\ast }$. Thus, the
first identity of fig. \ref{tad} holds.

The $L$-loop tadpole is equal to the $L$-th power of (\ref{umma}), divided
by $(-i\lambda )^{L-1}$. It satisfies the second identity of fig. \ref{tad},
because its marked version carries a minus sign for the marked vertex, while
the minus signs coming from the relation $iw=-iz^{\ast }$ still compensate
each other.

We have proved that, in the end, all Feynman diagrams satisfy the
diagrammatic cutting equations (\ref{mar}). From this point on, the proof of
perturbative unitarity can proceed according to the common strategy \cite%
{unitarity}.

The proof extends to (local) theories with derivative vertices and
propagators with nontrivial polynomial numerators. The algebraic cutting
equations are the same. The difference is that, before switching to the
diagrammatic cutting equations, we must multiply both sides of formula (\ref%
{theorem}) by appropriate polynomial numerators. We must show that the
right-hand side still vanishes after integrating on the energies. This is
less obvious than before.

When the numerator contains enough powers of the energy, contact terms may
appear. Contact terms collapse propagators and generate new types of
vertices and diagrams, which obey their own cutting equations. As shown in
ref. \cite{unitarity}, it is possible to associate each diagram $G$ with a
set of separate cutting equations that involve no contact terms, the sum of
which is equivalent to the $G$ cutting equations at $\epsilon \rightarrow 0$%
. For this reason, there is no loss of generality in assuming that contact
terms are absent. Since the propagators we are considering contain two
powers of the energy in the denominators, we can assume that the numerators
of the diagram $G$ contain at most one power of each loop energy.

As before, we can restrict to diagrams $G$ with no tadpoles, since tadpoles
are easily attached to $G$ at the end. Thus, every polarized loop $\gamma _{%
\text{pol}}$ that appears in $\mathcal{P}_{G}$ contains two or more internal
legs. If the internal legs are at least three, the energy integral is still
convergent: each polar number behaves like $1/E$ for large energy $E$, while
the numerator provides at most one $E$ power; since the integrand of a
polarized loop has all the poles on the same side of the integration path,
the residue theorem gives zero.

The only case that deserves attention is when the polarized loop $\gamma _{%
\text{pol}}$ has two legs, and, therefore, two vertices, which we call $\nu $
and $\nu ^{\prime }$. We can assume that $\gamma _{\text{pol}}$ is oriented.
Due to the nontrivial numerator, we get the integrals%
\begin{equation}
\frac{1}{4\omega _{1}\omega _{2}}\int_{-\infty }^{+\infty }\frac{\mathrm{d}E%
}{2\pi }\frac{E}{(E-\alpha \pm i\epsilon )(E-\beta \pm i\epsilon ^{\prime })}%
=\mp \frac{i}{8\omega _{1}\omega _{2}},  \label{resu}
\end{equation}%
depending on the polar numbers of $\gamma _{\text{pol}}$, where $\alpha $
and $\beta $ are real. We have not included the values of the vertices in
formula (\ref{resu}). We want to show that the contributions (\ref{resu})
cancel each other. The reason is that each polarized loop that contributes
with the upper sign, i.e. $\gamma _{\text{pol}}=\sigma _{1}^{+}\sigma
_{2}^{+}$, $\tau _{1}^{+}\sigma _{2}^{+}$, $\sigma _{1}^{+}\tau _{2}^{+}$, $%
\tau _{1}^{+}\tau _{2}^{+}$, is compensated by a polarized loop that
contributes with the lower sign, i.e. $\gamma _{\text{pol}}=\sigma
_{1}^{-}\sigma _{2}^{-}$, $\tau _{1}^{-}\sigma _{2}^{-}$, $\sigma
_{1}^{-}\tau _{2}^{-}$, $\tau _{1}^{-}\tau _{2}^{-}$.

Consider the left-hand side of equation (\ref{theorem}). Isolate the
contributions where $\gamma _{\text{pol}}=\sigma _{1}^{+}\sigma _{2}^{+}$
and $\gamma _{\text{pol}}=\sigma _{1}^{-}\sigma _{2}^{-}$. They come from
the diagrams $G_{M}$ where both $\nu $ and $\nu ^{\prime }$ are unmarked
(recall that $\gamma _{\text{pol}}$ is oriented) and the legs of the loop
are $z_{1}z_{2}$. Such diagrams compensate each other, when $\epsilon
\rightarrow 0$, because the coefficients of $\sigma _{1}^{+}\sigma _{2}^{+}$
and $\sigma _{1}^{-}\sigma _{2}^{-}$ are the same, but the values of the
polarized loops are opposite, by formula (\ref{resu}). A similar argument
applies to the pair $\tau _{1}^{+}\tau _{2}^{+}$ and $\tau _{1}^{-}\tau
_{2}^{-}$, which comes from $w_{1}w_{2}$ (with both $\nu $ and $\nu ^{\prime
}$ marked), as well as the pair $\sigma _{1}^{+}\tau _{2}^{+}$ and $\tau
_{1}^{-}\sigma _{2}^{-}$, which comes from $u_{1}v_{2}$, and finally the
pair $\tau _{1}^{+}\sigma _{2}^{+}$ and $\sigma _{1}^{-}\tau _{2}^{-}$,
which comes from $v_{1}u_{2}$. In the last two cases one vertex $\nu $ or $%
\nu ^{\prime }$ is marked and the other one is unmarked.

We conclude that the proof of perturbative unitarity based on the algebraic
cutting equations applies to all scalar field theories, including those that
have derivative vertices, as well as the nonrenormalizable ones. Following
the guidelines of ref. \cite{unitarity}, the proof can also be generalized
to the theories that include fermions, gauge fields and gravity, as long as
they are local, Hermitian and their kinetic terms are polynomials of degree
two (in the case of bosons) or degree one (in the case of fermions) in the
time derivatives.

\section{Parity symmetry}

\setcounter{equation}{0}\label{paritytr}

Some transformations relate algebraic cutting equations that may look
different, but are actually equivalent. Consider the polarity flipping, that
is to say the exchanges%
\begin{equation}
\sigma _{i}^{+}\longleftrightarrow \sigma _{i}^{-},\qquad \tau
_{i}^{+}\longleftrightarrow \tau _{i}^{-}.  \label{parity}
\end{equation}%
At the level of the propagators, this operation leaves $z_{i}$ and $w_{i}$
invariant and exchanges $u_{i}$ with $v_{i}$. By the Feynman rules of fig. %
\ref{feybub}, it is equivalent to flip the orientations of all the internal
legs of the diagram $G$. We call (\ref{parity}) \textit{parity}
transformation.

In the case of one-loop diagrams, the right-hand side $\mathcal{P}_{G}$ of
the identity (\ref{theorem}) is invariant, by formula (\ref{wou}).
Consequently, the left-hand side is also invariant. However, this fact may
become apparent only after expanding it as a sum of polar monomials. Check
for example the identity (\ref{polo}), associated with the triangle diagram.

When the number of loops exceeds one, both the left- and right-hand sides of
(\ref{theorem}) may change under the parity transformation. For example, it
is easy to check that $\mathcal{P}_{G}$ does change in the case of the
self-energies of fig. \ref{polloop}.

The diagrams $G$\ and $\bar{G}$ and so the left-hand side of the
diagrammatic cutting equation (\ref{mar}) are invariant. For this reason, (%
\ref{parity}) is a symmetry of the diagrammatic cutting equations.
Nevertheless, the right-hand side of (\ref{mar}) may get organized
differently after the transformation.

The algebraic cutting equations likely possess other hidden symmetries that
are awaiting to be uncovered. For example, the parity transformation (\ref%
{parity}) can be performed on just one or more legs, instead of all of them.
Moreover, the cut and uncut propagators are in some sense dual to each
other, because both are linear combinations of half propagators and their
different roles only emerge at the graphical level.

\section{Conclusions}

\setcounter{equation}{0}\label{conclusions}

In this paper, we have proved a set of algebraic identities that provide a
clearer understanding of perturbative unitarity in quantum field theory. To
conclude, we make some remarks on the virtues of the algebraic approach to
perturbative unitarity, in comparison with the usual approach.

When tadpoles are absent, equation (\ref{mar}) holds for arbitrary positive
values of the widths $\epsilon $ of the propagators (\ref{compl}). In
particular, the widths $\epsilon $ need not be infinitesimal. If we choose a
different $\epsilon $ for each internal leg, our algebraic theorem allows us
to keep track of them efficiently throughout the calculation. Each
propagator (cut or not) keeps its own $\epsilon $ from the beginning to the
end and no mixing between the $\epsilon $s of different propagators does
occur. This means that we are allowed to freely send them to zero in the
order we want. When we do it, the cut propagators become those we are
accustomed to, i.e. 
\begin{equation*}
\lim_{\epsilon \rightarrow 0}iu_{\mathbf{k}\epsilon }=(2\pi )\theta
(k^{0})\delta (k^{2}-m^{2}),\qquad \lim_{\epsilon \rightarrow 0}iv_{\mathbf{k%
}\epsilon }=(2\pi )\theta (-k^{0})\delta (k^{2}-m^{2}).
\end{equation*}%
Yet, we stress again that the identity (\ref{mar}) also holds when the cut
propagators are $iu_{\mathbf{k}\epsilon }$ and $iv_{\mathbf{k}\epsilon }$,
where $\epsilon $ is arbitrary, at least when tadpoles are absent. When
tadpoles are present, the widths can be arbitrary everywhere but in the
tadpoles, where they must be set to zero to ensure that the tadpole
assumption holds. Similar arguments hold for the contact terms, which lead
to formulas such as (\ref{resu}), when derivative vertices are present.

If we do not use the algebraic theorem of this paper and make rather natural
operations on the integrands, it is easy to generate inconvenient mixings
between the $\epsilon $s of different propagators and encounter ill-defined
distributions such as \cite{piva}%
\begin{equation}
\frac{1}{\omega -\omega ^{\prime }-p^{0}-i(\epsilon _{1}-\epsilon _{2})},
\label{ill}
\end{equation}%
where $\omega =\sqrt{\mathbf{k}^{2}+m_{1}^{2}}$ and $\omega ^{\prime }=\sqrt{%
(\mathbf{k}-\mathbf{p})^{2}+m_{2}^{2}}$ are some frequencies, $p$ is an
external momentum and $k$ is a loop momentum. It is possible to show (see
ref. \cite{piva} for details) that the ill-defined part of (\ref{ill})
ultimately does not contribute. The theorem proved here guides us through
the calculations without ever meeting these ill-defined distributions.

Going through the analysis recently made in ref. \cite{unitarity}, where the
assumptions behind the proof of perturbative unitarity have been relaxed to
a minimum, it is possible to realize that the properties just emphasized can
also be proved in the usual nonalgebraic approach. However, an approach like
the algebraic one, which makes them so apparent, is of great advantage.

The usual approach is also responsible for giving some false impressions.
For example, it suggests that the cut propagators must force the energy
propagation in a given direction. This is not true, as the validity of (\ref%
{mar}) at arbitrary, nonvanishing widths points out. Again, it is not
impossible to show this fact in the usual approach, because the assumption
about the energy flow enters the proof only at a later stage \cite{unitarity}%
. However, the roles of the various ingredients of the proof become much
clearer when the algebraic cutting equations are used.

The algebraic approach is useful to prove perturbative unitarity\ to all
orders in theories that have not been reached by more standard techniques,
as recently shown in ref. \cite{fakeons} for the Lee-Wick models and the
fakeon models.

\vskip 22truept \noindent {\large \textbf{Acknowledgments}}

\vskip 2truept

We are grateful to U. Aglietti and M. Piva for useful discussions.


\begin{thebibliography}{99}
\bibitem{cutkosky} R.E. Cutkosky, Singularities and discontinuities of
Feynman amplitudes, \href{https://doi.org/10.1063/1.1703676}{J. Math. Phys.
1 (1960) 429};

M. Veltman, Unitarity and causality in a renormalizable field theory with
unstable particles, \href{https://doi.org/10.1016/S0031-8914(63)80277-3}{%
Physica 29 (1963) 186}.

\bibitem{thooft} G. 't Hooft, Renormalization of massless Yang-Mills fields, 
\href{https://doi.org/10.1016/0550-3213(71)90395-6}{Nucl.Phys. B 33 (1971)
173};

G. 't Hooft, Renormalizable Lagrangians for massive Yang-Mills fields, \href{https://doi.org/10.1016/0550-3213(71)90139-8}%
{Nucl. Phys. B} \href{https://doi.org/10.1016/0550-3213(71)90139-8}%
{35 (1971) 167}.

\bibitem{brsunitarity} %C. Becchi, A. Rouet and R. Stora, The Abelian Higgs
%Kibble model, unitarity of the S-operator, Phys. Lett. B 52 (1974) 344;

%C. Becchi, A. Rouet and R. Stora, Renormalization of gauge theories, Ann.
%Phys. (NY) 98 (1976) 287;

See for example, G. Curci and R. Ferrari, An alternative approach to the
proof of unitarity for gauge theories, \href{https://doi.org/10.1007/BF02730284}%
{Nuovo Cimento A 35 (1976) 273}, and references therein.

%T. Kugo and I. Ojima, Local covariant operator formalism of non-Abelian
%gauge theories and quark confinement problem, Prog. Theor. Phys. Suppl. 66
%(1979) 1;

%C. Becchi, \textquotedblleft Lectures on the renormalization of gauge
%theories,\textquotedblright\ in Les Houches 1983, Proceedings, Relativity,
%Groups and Topology, II, p. 787.

\bibitem{unitarity} D. Anselmi, Aspects of perturbative unitarity, \href{http://dx.doi.org/10.1103/PhysRevD.94.025028}%
{Phys. Rev. D 94 (2016) 025028}, \href{http://renormalization.com/16a1/}{%
16A1 Renormalization.com} and \href{http://arxiv.org/abs/1606.06348}{%
arXiv:1606.06348} [hep-th].

\bibitem{diagrammar} G. 't Hooft and M.J. Veltman, \textit{Diagrammar},
report No. CERN-73-09, available at \href{http://cds.cern.ch/record/186259/files/p1.pdf}%
{this link.}

\bibitem{leewick} T.D. Lee and G.C. Wick, Negative metric and the unitarity
of the S-matrix, \href{https://doi.org/10.1016/0550-3213(69)90098-4}{Nucl.
Phys. B 9 (1969) 209};

T.D. Lee and G.C. Wick, Finite theory of quantum electrodynamics, \href{https://doi.org/10.1103/PhysRevD.2.1033}%
{Phys. Rev. D 2 (1970) 1033};

R.E. Cutkosky, P.V Landshoff, D.I. Olive, J.C. Polkinghorne, A non-analytic
S matrix, \href{https://doi.org/10.1016/0550-3213(69)90169-2}{Nucl. Phys.
B12 (1969) 281}.

\bibitem{contexts} B. Grinstein, D. O'Connell, and M.B. Wise, The Lee-Wick
standard model, \href{https://doi.org/10.1103/PhysRevD.77.025012}{Phys. Rev.
D77 (2008) 025012} and \href{http://arxiv.org/abs/0704.1845}{arXiv:0704.1845}
[hep-ph];

C.D. Carone and R.F. Lebed, Minimal Lee-Wick extension of the standard
model, \href{https://doi.org/10.1016/j.physletb.2008.08.050}{Phys. Lett.
B668 (2008) 221} and \href{http://arxiv.org/abs/0806.4555}{arXiv:0806.4555}
[hep-ph];

J.R. Espinosa and B. Grinstein, Ultraviolet properties of the Higgs sector
in the Lee-Wick standard model, \href{https://doi.org/10.1103/PhysRevD.83.075019}%
{Phys. Rev. D83 (2011) 075019} and \href{http://arxiv.org/abs/1101.5538}{%
arXiv:1101.5538} [hep-ph];

C.D. Carone and R.F. Lebed, A higher-derivative Lee-Wick standard model, 
\href{https://doi.org/10.1088/1126-6708/2009/01/043}{J. High Energy Phys.
0901 (2009) 043} and \href{http://arxiv.org/abs/0811.4150}{arXiv:0811.4150}
[hep-ph];

B. Grinstein and D. O'Connell, One-Loop Renormalization of Lee-Wick Gauge
Theory, \href{https://doi.org/10.1103/PhysRevD.78.105005}{Phys. Rev. D78
(2008) 105005} and \href{http://arxiv.org/abs/0801.4034}{arXiv:0801.4034}
[hep-ph];

C. D. Carone, Higher-derivative Lee-Wick unification, \href{https://doi.org/10.1016/j.physletb.2009.05.053}%
{Phys. Lett. B677 (2009) 306} and \href{http://arxiv.org/abs/0904.2359}{%
arXiv:0904.2359} [hep-ph].

\bibitem{LWgrav} E. Tomboulis, 1/N expansion and renormalization in quantum
gravity, Phys. \href{https://doi.org/10.1016/0370-2693(77)90678-5}{Lett. B
70 (1977) 361};

E. Tomboulis, Renormalizability and asymptotic freedom in quantum gravity, 
\href{https://doi.org/10.1016/0370-2693(80)90550-X}{Phys. Lett. B 97 (1980)
77};

L. Modesto, Super-renormalizable or finite Lee--Wick quantum gravity, \href{https://doi.org/10.1016/j.nuclphysb.2016.06.004}%
{Nucl. Phys. B909 (2016) 584} and \href{http://arxiv.org/abs/1602.02421}{%
arXiv:1602.02421} [hep-th];

I. Shapiro and L. Modesto, Superrenormalizable quantum gravity with complex
ghosts, \href{https://doi.org/10.1016/j.physletb.2016.02.021}{Phys. Lett.
B755 (2016) 279} and \href{http://arxiv.org/abs/1512.07600}{arXiv:1512.07600}
[hep-th].

\bibitem{LWgravmio} D. Anselmi, On the quantum field theory of the
gravitational interactions, \href{http://dx.doi.org/doi:10.1007/JHEP06(2017)086}%
{J. High Energy Phys. 06 (2017) 086}, \href{http://renormalization.com/17a3/}%
{17A3 Renormalization.com} and \href{http://arxiv.org/abs/1704.07728}{%
arXiv:1704.07728} [hep-th].

\bibitem{LWformulation} D. Anselmi and M. Piva, A new formulation of
Lee-Wick quantum field theory, \href{http://dx.doi.org/10.1007/JHEP06(2017)066}%
{J. High Energy Phys. 06 (2017) 066}, \href{http://renormalization.com/17a1/}%
{17A1 Renormalization.com} and \href{http://arxiv.org/abs/1703.04584}{%
arXiv:1703.04584} [hep-th].

\bibitem{piva} D. Anselmi and M. Piva, Perturbative unitarity in Lee-Wick
quantum field theory, \href{http://dx.doi.org/10.1103/PhysRevD.96.045009}{%
Phys. Rev. D 96 (2017) 045009} and \href{http://renormalization.com/17a2/}{%
17A2 Renormalization.com} and \href{http://arxiv.org/abs/1703.05563}{%
arXiv:1703.05563} [hep-th].

\bibitem{fakeons} D. Anselmi, Fakeons and Lee-Wick models, \href{http://dx.doi.org/10.1007/JHEP02(2018)141}%
{J. High Energy Phys. 02 (2018)} 141, \href{http://renormalization.com/18a1/}%
{18A1 Renormalization.com} and \href{http://arxiv.org/abs/1801.00915}{%
arXiv:1801.00915} [hep-th].
\end{thebibliography}
\end{document}